\documentclass[12pt,oneside]{article}
\usepackage{geometry}
\usepackage[pdftex]{graphicx}
\usepackage{amsmath}
\usepackage{amssymb}
\usepackage{amsfonts}
\usepackage[usenames,dvipsnames]{xcolor}
\usepackage[utf8]{inputenc}
\usepackage{multicol}
\usepackage{multirow}
\usepackage{verbatim}
\usepackage[small]{caption}
\usepackage{rotating}
\usepackage{float}
\usepackage{url}
\usepackage{pdflscape}
\usepackage{enumerate}
\usepackage{footnote}
\makesavenoteenv{tabular}
\usepackage[T1]{fontenc}
\usepackage{lmodern}
\usepackage[sort&compress,numbers]{natbib}
\usepackage[export]{adjustbox}
\usepackage[onehalfspacing]{setspace}
\usepackage{threeparttable}
\usepackage{afterpage}
\usepackage[normalem]{ulem}
\usepackage{color}
\usepackage{mwe,tikz}\usepackage[percent]{overpic}
\usepackage[onehalfspacing]{setspace}
\usepackage{hyperref}

\title{\textbf{Ship-track-based assessments overestimate the cooling effect of anthropogenic aerosol}\thanks{
This preprint is intended for publication in a scientific journal but has not been peer-reviewed.
The copyright is maintained by the authors or by other copyright owners.
It is understood that all persons copying this information will adhere to the terms and constraints invoked by this copyright.}} 

\author{Franziska Glassmeier,$^{1,2,3\star}$ Fabian Hoffmann,$^{3,4}$ Jill S. Johnson,$^{5}$ \\ Takanobu Yamaguchi,$^{3,4}$ Ken S. Carslaw,$^{5}$ Graham Feingold$^{4}$\\[18pt]
\footnotesize{$^{1}$Department Geoscience and Remote Sensing, Delft University of Technology, Netherlands}\\[-9pt]
\footnotesize{$^{2}$Department of Environmental Sciences, Wageningen University, Netherlands}\\[-9pt]
\footnotesize{$^{3}$Cooperative Institute for Research in Environmental Sciences, University of Colorado, USA}\\[-9pt]
\footnotesize{$^{4}$NOAA Chemical Sciences Laboratory, Boulder, USA}\\[-9pt]
\footnotesize{$^{5}$School of Earth and Environment, University of Leeds, UK}\\[18pt]
\normalsize{$^\star$Correspondence: f.glassmeier@tudelft.nl}
}

\date{}

\begin{document} 
\maketitle 
\vspace{-24pt}
\thispagestyle{empty}
  \begin{singlespace}
  \begin{abstract}
  \noindent The effect of anthropogenic aerosol on the reflectivity of stratocumulus cloud decks through changes in cloud amount is a major uncertainty in climate projections.
  The focus of this study is the frequently occurring non-precipitating stratocumulus.
  In this regime, cloud amount can decrease through aerosol-enhanced cloud-top mixing.
  The climatological relevance of this effect is debated
  because ship exhaust does not appear to generate significant change in the amount of these clouds.
  Through a novel analysis of detailed numerical simulations in comparison to satellite data, we show that results from ship-track studies cannot be generalized to estimate the climatological forcing of anthropogenic aerosol.
  We specifically find that the ship-track-derived sensitivity of the radiative effect of non-precipitating stratocumulus to aerosol overestimates their cooling effect by up to 200\,\%.
    This offsetting warming effect needs to be taken into account if we are to constrain the aerosol-cloud radiative forcing of stratocumulus.
  \end{abstract}
  \end{singlespace}

\newpage
\setcounter{page}{1}
\subsubsection*{Introduction}

Clouds interact with radiation and therefore play an important role in the planetary energy balance.
Their net effect is to cool the planet by reflecting incoming short-wave radiation \cite{Stephens_2012}.
Covering large parts of the sub-tropical oceans, stratocumulus (Sc) clouds are by far the largest contributor to this cooling \cite{L_Ecuyer_2019}.
Anthropogenic perturbations to cloud reflectivity that result from an increased concentration of atmospheric aerosol particles are the most uncertain anthropogenic forcing of the climate system \cite{BoucherRandall13, Bellouin_2019}.
As a striking illustration of this effect, exhaust from ships can create ``ship tracks'' that manifest as bright linear features in Sc decks.
This brightening arises because exhaust-aerosol particles form the nuclei of cloud droplets.
A greater abundance of particles means that a cloud consists of more, but smaller droplets, which enhances the radiant energy reflected to space \cite{Twomey74}.
Changes in the number and size of cloud droplets also influence cloud physical processes \cite{Albrecht89, WangWangFeingold03, Ackermann_2004, BrethertonBlosseyUchida07, SmallChuangFeingold09, Hoffmann_2019};
for the example of ship tracks this means that the amount of cloud water inside and outside of a track may evolve differently.
Globally, the large uncertainty in the cloud-mediated aerosol forcing arises from the unknown magnitude of such \emph{adjustments} of cloud water in response to aerosol-induced perturbations \cite{StevensFeingold09, BoucherRandall13, MuelmenstaedtFeingold18}.
Here we show that despite providing a striking illustration of aerosol-cloud interactions, ship tracks {\em do not} provide suitable data to estimate the magnitude of cloud liquid-water adjustments in a polluted climate, in contrast
with the common assumption that ship tracks {\em do} represent such adjustments \cite{ChristensenStephens11, GryspeerdtGorenSourdeval19, TollChristensenQuaas19, DiamondDirectorEastman20}.

\begin{figure}[t]
  \centering
  \begin{overpic}[width=0.8\textwidth]{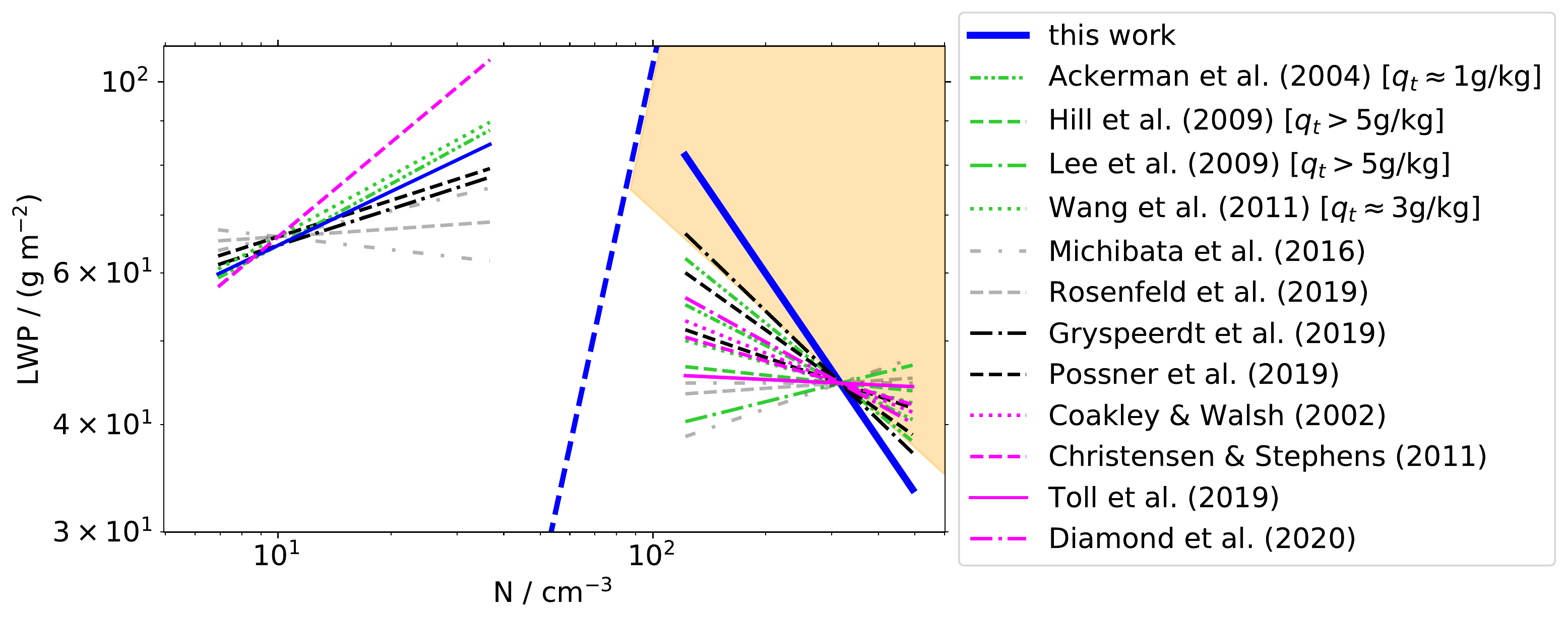}
    \put(12, 8){\tiny\textcolor{blue}{precipitating}}
    \put(38, 9){\tiny\textcolor{blue}{non-precipitating/}}
    \put(38, 7){\tiny\textcolor{blue}{entrainment-dominated}}
    \put(45, 33){\footnotesize\textcolor{orange}{warming}}
  \end{overpic}
  \caption{Reported log-log-linear relationships between liquid-water path, LWP, and cloud droplet number, $N$, in comparison to this work.
  Lines are based on reported slopes (Table~\ref{tab:lit}),
  axis intercepts have been added as suitable for illustration.
  The dashed blue line indicates a critical droplet radius for precipitation formation
  (based on a mean droplet radius of $12\,\mu$m at cloud top for an adiabatic condensation rate of $2.5 \cdot 10^{-6}\,\textrm{kg}\,\textrm{m}^{-4}$),
  which separates the precipitation-dominated regime on the left from the non-precipitating, entrainment-dominated regime on the right.
  The latter is the focus of this study.
  Colors distinguish results from large-eddy simulations (green), climatological satellite studies (black), and satellite studies of ship tracks (magenta).
  Results in grey are shown for completeness but are not directly comparable due to differences in methodology.
  For large-eddy simulation studies, above-cloud absolute humidity $q_t$ is indicated.
  Solid blue lines show values derived in this work (Table~\ref{tab:lit} and~\ref{tab:uncertainty}, in particular $\textrm{d}\ln\textrm{LWP}/\textrm{d} \ln N = -0.64$ in the entrainment regime).
  The orange shading indicates where LWP adjustments are sufficiently negative to lead to climate warming rather than cooling based on the sign of albedo sensitivity $S$ (Equation~\ref{eq:sensitivity}).
    \label{fig:lit}}
\end{figure}
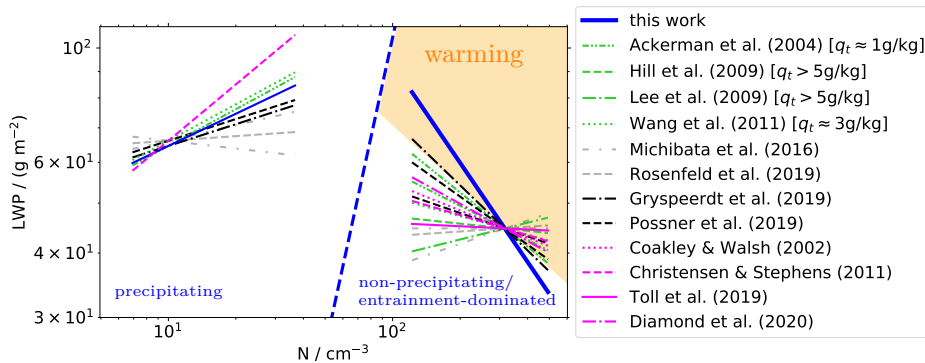

In non-precipitating Sc, cloud response to aerosol perturbations is commonly quantified by the sensitivity  \cite{Bellouin_2019, PlatnickTwomey94, Boers_1994}
\begin{equation}
S = \frac{\textrm{d}A_\textrm{c}}{\textrm{d}N}
  = \frac{A_\textrm{c}\left(1-A_\textrm{c}\right)}{3N} \left(1 + \frac{5}{2} \frac{\textrm{d}\ln \textrm{LWP}}{\textrm{d}\ln N} \right)
  \label{eq:sensitivity}
\end{equation}
of cloud albedo $A_\textrm{c}$ to cloud droplet number $N$.
The first term on the right-hand side of Equation~\ref{eq:sensitivity} quantifies the albedo effect of changing droplet number when keeping the vertically integrated amount of liquid water, or \emph{liquid-water path}, $\textrm{LWP}$, constant;
the second term accounts for cloud water adjustments as quantified by the relative sensitivity $\textrm{d}\ln\textrm{LWP}/\textrm{d}\ln N$ of $\textrm{LWP}$ to $N$.
Numerical values for LWP-adjustments $\textrm{d}\ln\textrm{LWP}/\textrm{d}\ln N$ have been derived from detailed modeling and satellite studies \cite{CoakleyWalsh02, Ackermann_2004, Hill_2009, LeePennerSaleeby09, Wang_2011, ChristensenStephens11, Michibata_2016, GryspeerdtGorenSourdeval19, Rosenfeld_2019, PossnerEastmanBender19, TollChristensenQuaas19, DiamondDirectorEastman20}.
Both approaches have recently converged on the insight that the sign of LWP adjustments is regime-dependent (Figure~\ref{fig:lit}).
Adjustments tend to be positive under precipitating conditions where the addition of particles decreases drop size, increases colloidal stability, and allows for an accumulation of liquid water \cite{Albrecht89}.
A positive LWP adjustment thus implies thicker, more reflective clouds that have a stronger cooling effect.
The effects of aerosol perturbations on precipitation were considered recently \cite{Rosenfeld_2019}.
In the current work we focus on non-precipitating Sc whose development is dominated by entrainment.
Observations show that this Sc regime is more common than the precipitating regime \cite{LeonWangLiu08, PossnerEastmanBender19}.
Non-precipitating Sc feature negative adjustments, indicating a decrease in LWP for higher aerosol concentrations.
The decrease in LWP stems from the accelerated and stronger evaporation of cloud liquid in higher aerosol conditions as the Sc mixes with dry air from above the cloud (\emph{entrainment}).
Smaller droplets evaporate more efficiently because they provide a larger surface (for a given total amount of liquid)
and reside closer to the entrainment interface than larger droplets due to reduced gravitational settling, which increases the potential for evaporation \cite{WangWangFeingold03, Ackermann_2004, BrethertonBlosseyUchida07, XueFeingold2008, SmallChuangFeingold09, Hoffmann_2019}.
Negative LWP adjustment values indicate thinner, less reflective clouds and a weaker cooling effect.
When the darkening effect of cloud thinning is stronger than the brightening of increased $N$,
negative LWP adjustments can even imply a warming effect. 
In non-precipitating Sc, this is the case when $\textrm{d}\ln \textrm{LWP}/\textrm{d}\ln N <-2/5$ such that Equation~\ref{eq:sensitivity} becomes negative (orange shading in Figure~\ref{fig:lit}).

In addition to the distinction between the entrainment- and precipitation-dominated regimes, satellite studies have identified above-cloud moisture as an important control on the magnitude of LWP adjustments in Sc \cite{
ChenChristensenStephens14,
Michibata_2016,
GryspeerdtGorenSourdeval19,
PossnerEastmanBender19}.
This is consistent with process-understanding from detailed cloud modeling studies (\emph{large-eddy simulation}, LES), where drier above-cloud conditions correspond to a stronger aerosol-effect on entrainment (Figure~\ref{fig:lit}).
As another factor behind the variability of adjustment estimates, references~\cite{GryspeerdtGorenSourdeval19, BenderFreyMcCoy19} discuss the effects of \emph{$N$-\emph{LWP} co-variability} that results from large-scale co-variability of aerosol and moisture.
As an example of this confounding effect, compare a maritime situation with a clean and moist atmosphere to a polluted and drier continental case.
Observations from these two cases will likely show that higher $N$ is correlated with lower LWP \cite{BrenguierPawlowskaSchuller03}, suggesting a negative LWP-adjustment.
Clearly, the ``adjustment'' quantified here is not related to the effect of aerosol on cloud properties driven by entrainment or precipitation formation that we seek to capture, but rather, to large-scale conditions.

A special appeal of ship tracks has been that they are not affected by external co-variability because the large-scale meteorological conditions are the same inside and outside of the track.
Accordingly, results from targeted satellite analyses of ship-tracks \cite{CoakleyWalsh02, ChristensenStephens11, TollChristensenQuaas19} have been assigned higher credibility than climatological satellite studies, for which external co-variability cannot be ruled out.
In particular, the comparably large absolute adjustment values found in the latter studies have been attributed to aerosol-moisture co-variability, assuming that weak-to-almost absent LWP adjustments identified by ship-track studies \cite{CoakleyWalsh02, ChristensenStephens11, TollChristensenQuaas19} provide the best estimate for LWP adjustment.
In contrast to this assumption, a recent study of shipping lanes reports significantly negative adjustment values \cite{DiamondDirectorEastman20}.

\begin{figure}
  \begin{overpic}[width=\textwidth]{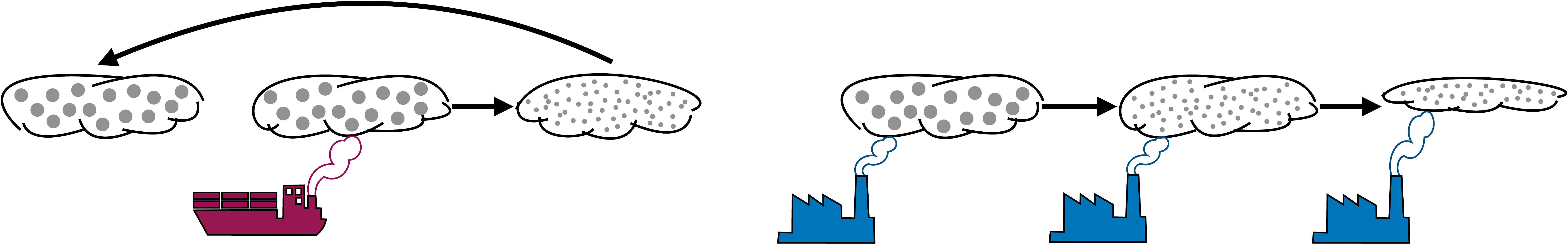}
    \put(29, 10.5){(1)}
    \put(20, 17){(2a)}
    \put(67, 10.5){(1)}
    \put(83.5, 10.5){(2b)}
  \end{overpic}
  \caption{LWP adjustments in ship tracks, which persist for a few hours, as compared to industrial-era pollution, which perturbs the climatological aerosol background and leads to perturbations that last for days.
  As an initial response to the aerosol perturbation, both situations feature cloud brightening through more but smaller cloud droplets at constant LWP (step 1).
  The ship track then returns to its original state because the perturbation ceases (step 2a).
  An enhanced aerosol background, in contrast, persists and allows for LWP to equilibrate to a new steady state that is characterized by increased entrainment efficiency and a lower LWP (step 2b).
  \label{fig:cartoon}}
\end{figure}
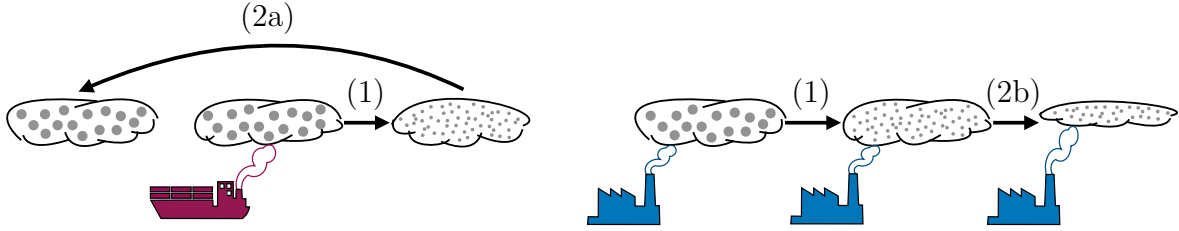

In this article, we show that the current emphasis on satellite studies of ship tracks to estimate LWP adjustments leads to an overestimation of the cooling effect of aerosols in Sc. 
We furthermore reconcile the broad range of reported adjustment estimates and discuss implications of our results for identifying alternatives to ship-track studies. 
Our argument is illustrated in Figure~\ref{fig:cartoon} and builds on two key results:
Firstly, LWP adjustments become more negative as Sc decks evolve towards a steady-state, bounded from below by $\textrm{d}\ln \textrm{LWP}/\textrm{d}\ln N = -0.64$.  
Secondly, in ship tracks, this temporal evolution does not proceed long enough to be representative of Sc decks in a polluted climate.

\subsubsection*{Data and methods}

\begin{figure}[t]
  (a) \hspace{0.49\textwidth} (b)\\
  \begin{overpic}[width=0.49\textwidth]{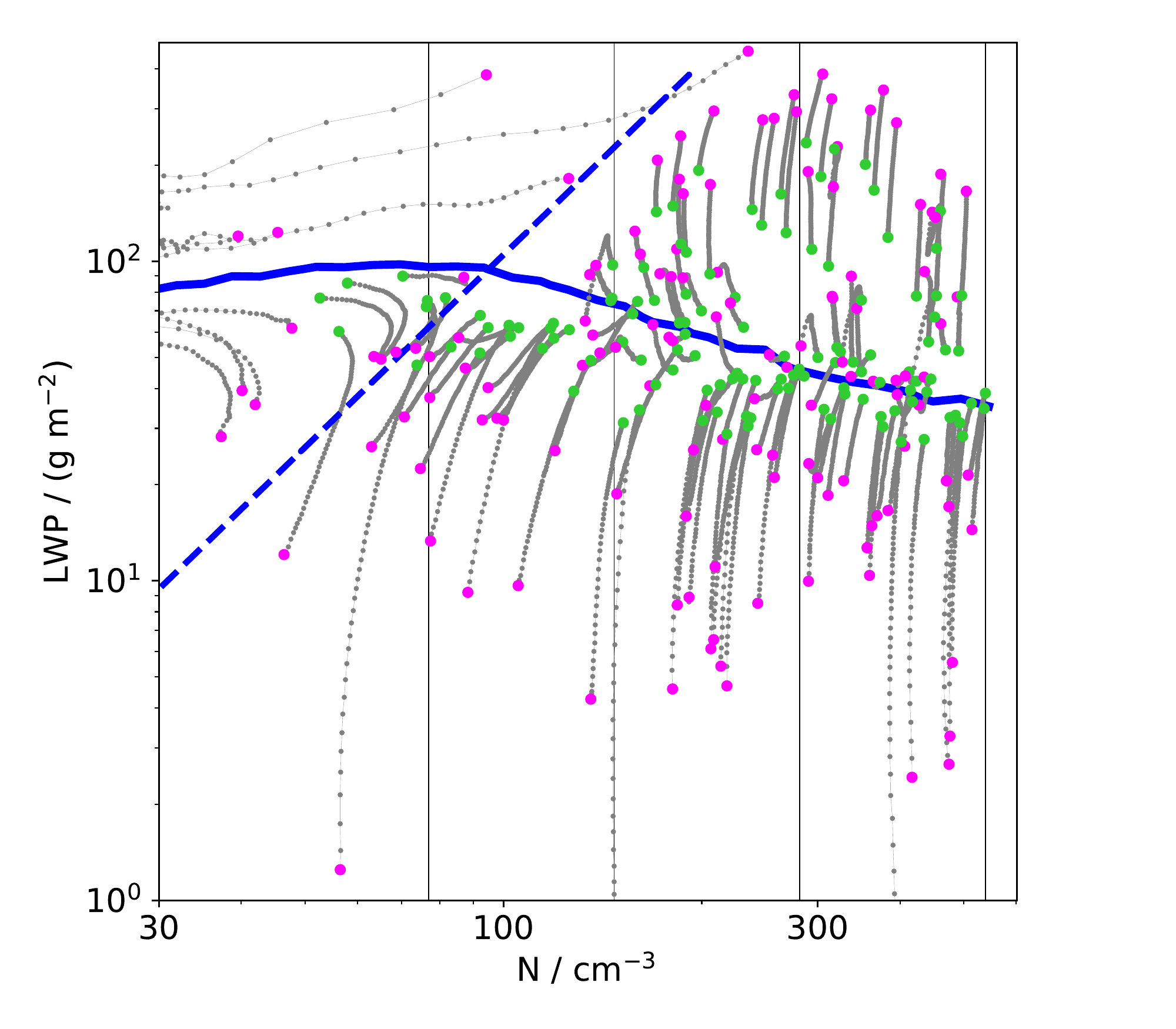}
    \put(15,35){\colorbox{gray!30}{\textcolor{blue}{\tiny $r_\textrm{crt}=12\,\mu$m}}}
    \put(15,68){\colorbox{gray!30}{\textcolor{blue}{\tiny steady-state LWP}}}
  \end{overpic}
  \hfill
  \begin{overpic}[width=0.49\textwidth]{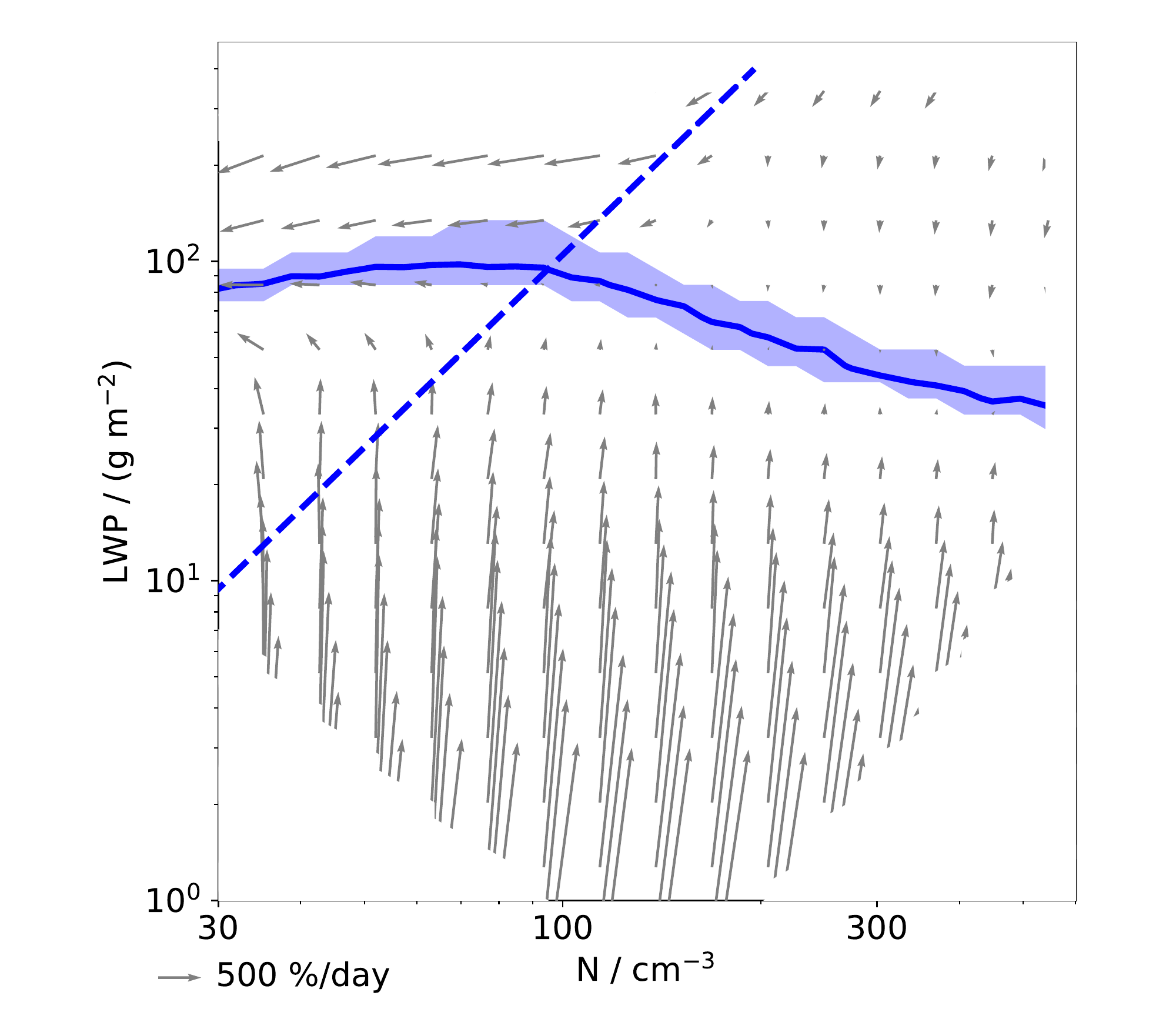}
    \put(20,35){\colorbox{gray!30}{\textcolor{blue}{\tiny $r_\textrm{crt}=12\,\mu$m}}}
    \put(20,68){\colorbox{gray!30}{\textcolor{blue}{\tiny steady-state LWP}}}
  \end{overpic}
  \caption{LES ensemble dataset and corresponding temporal co-evolution of liquid-water path LWP and cloud-droplet number $N$, focusing on the non-precipitating regime below the dashed blue line indicating the critical radius for precipitation formation as in Figure~\ref{fig:lit}.
  (a) Temporal evolution of 144 LES runs with varying initial conditions. 
  Individual simulation runs are indicated by gray lines connecting gray circles.
  Start (2\,h into the simulation to allow for model spin-up, magenta) and end (12\,h, green) of a trajectory are color-highlighted.
  The solid blue line shows the steady-state LWP from (b).
  Thin vertical black lines indicate the boundaries of $N$-bins used in Figure~\ref{fig:adj_fits}.
  (b) Flow-field representation of LWP-$N$ co-evolution and location of steady-state LWP (blue line and 25th/75th percentile uncertainty shading, Table~\ref{tab:uncertainty}), which is characterized by $\textrm{d}\ln\textrm{LWP}/\textrm{d}t = 0$.
  \label{fig:methods}}
\end{figure}

Our analysis builds on relating satellite datasets to LES.
We make these two data sources comparable by creating an ensemble of 144 LES runs that resembles the scope of a satellite dataset in that it samples a broad range of LWP and $N$ conditions (Figure~\ref{fig:methods}, \emph{Supplementary Information}, \cite{GlassmeierHoffmannJohnson19, FeingoldMcComiskeyYamaguchi16}).
In contrast to satellite data, we prevent externally-induced $N$-LWP co-variability by sampling initial conditions of ensemble members in a statistically independent way (\emph{Supplementary Information}, \cite{FeingoldMcComiskeyYamaguchi16}).
We furthermore limit externally-induced variability in LWP adjustments by fixing external control parameters of Sc (Table~\ref{tab:sim_parameters}).
We specifically restrict above-cloud absolute humidity to values $q_\textrm{t} < 2.8\,\textrm{g}/\textrm{kg}$ with a median value of $0.5\,\textrm{g}/\textrm{kg}$.
This choice of very dry above-cloud conditions allows us to derive a lower bound for LWP adjustments.

We approach our investigation of LWP adjustments to $N$ perturbations by analyzing the temporal co-evolution of LWP and $N$ collectively for all members of the Sc ensemble (Figure~\ref{fig:methods}). 
In the LWP direction, individual ensemble members collectively evolve towards similar LWPs (along an approximately horizontal line in Figure~\ref{fig:methods}).
For these steady-state LWPs, there exists a balance in the contributions of different processes that are source and sink terms for LWP - in particular radiative cooling (source), and entrainment and precipitation drying (sink) \cite{HoffmannGlassmeierFeingold19}.
The collective evolution in the $N$ direction is structured around the critical radius for precipitation formation \cite{RosenfeldGutman94}.
Precipitating systems (above the dashed line in Figure~\ref{fig:methods}) contain sufficient drops with radii larger than the critical radius, are colloidally unstable, and feature a rapid reduction in $N$.
For systems with smaller radii (below the dashed line), which are our focus, rain is scant and entrainment dominates.

The individual evolution of Sc cloud fields in the ensemble can be represented as flow vectors $\vec{v} = \left(\textrm{d}\ln N/\textrm{d}t, \textrm{d} \ln \textrm{LWP} /\textrm{d} t\right)^\textrm{T}$ in $N$-LWP space.
\emph{Gaussian-process emulation} allows us to interpolate such flow vectors from our limited number of simulations (Figure~\ref{fig:methods}, a) to obtain the full flow field illustrated in Figure~\ref{fig:methods} (b) (\emph{Supplementary Information}). 
This interpolation of the flow field enables us to infer cloud behavior beyond the 12h-duration of our simulations, including the behavior when an LWP \emph{steady state} ($\textrm{d}\ln\textrm{LWP}/\textrm{d}t = 0$) is reached (blue curve in Figure~\ref{fig:methods}, b).
This enables us to systematically quantify the time-dependence of LWP adjustments over timescales longer than the duration of our simulations.
The flow field representation also allows us to determine that individual Sc systems equilibrate to their steady-state with a characteristic equilibration time scale of $\tau=9.6\,$h (Figure~\ref{fig:ts}, \cite{Strogatz94}), in excellent agreement with a theoretical estimate \cite{SchubertWakefieldSteiner79}.
This timescale informs us about the proximity of an observed Sc system to its steady state.

\subsubsection*{Effect of cloud field evolution towards steady state on adjustment strength}

\begin{figure}[t]
  (a) \hspace{0.435\textwidth} (b)\\
  \includegraphics[height=0.39\textwidth]{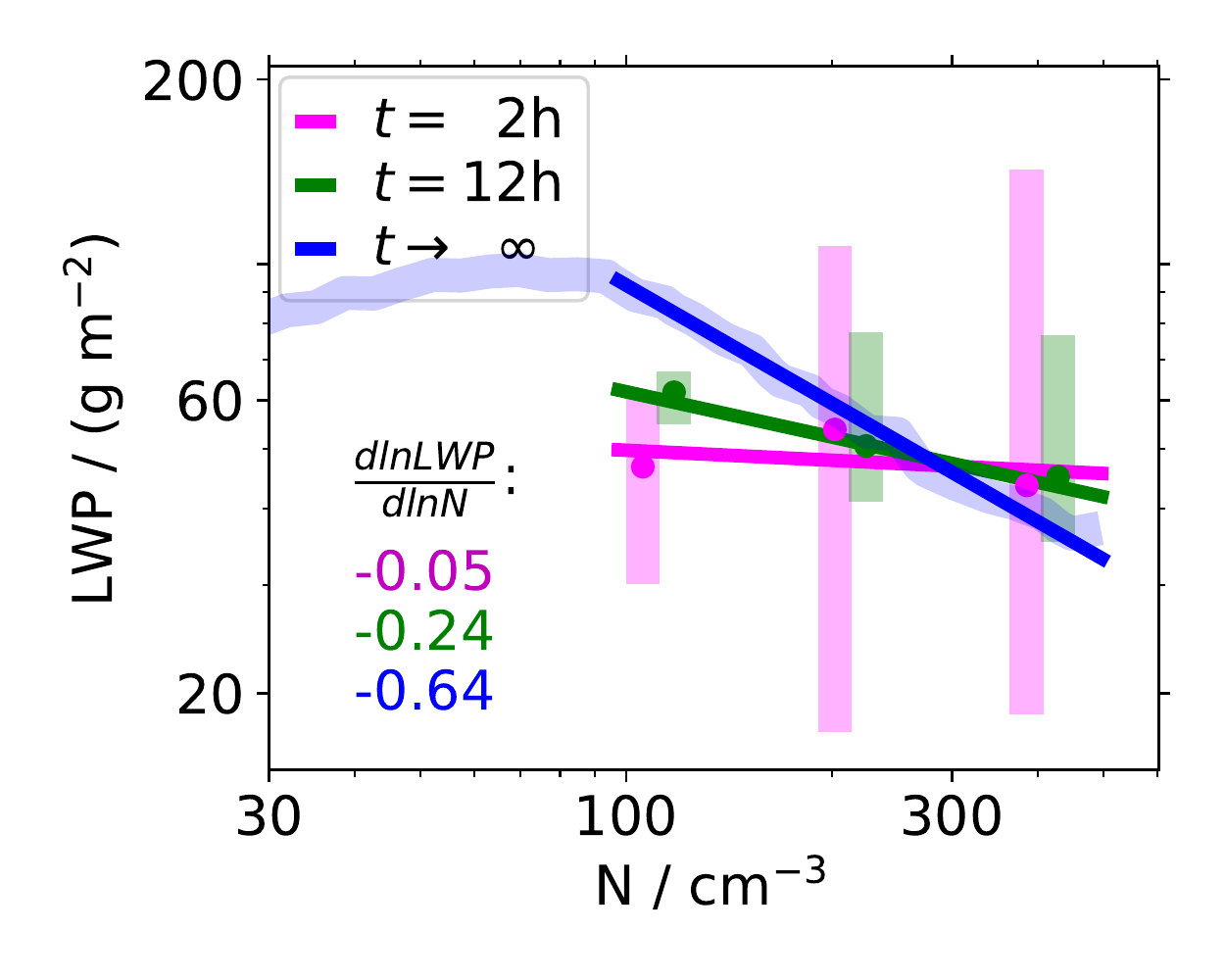}
  \includegraphics[height=0.39\textwidth]{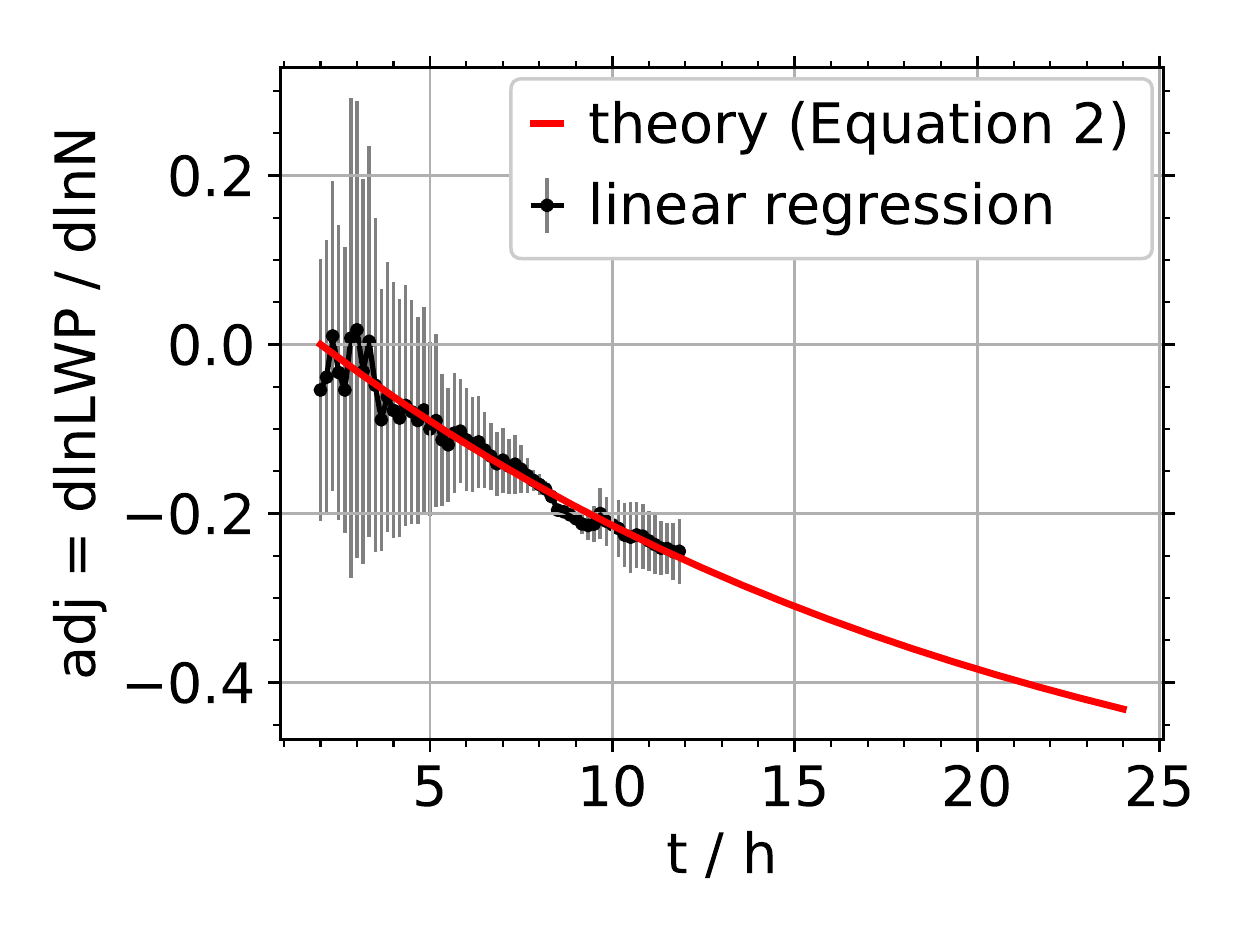}
  \caption{Time-dependence of LWP-adjustments.
  (a) Data points with error bars show median and 25th/75th percentile of simulated LWP at (magenta) $t=2\,$h and (green) $t=12\,$h for the $N$-bins indicated in Figure~\ref{fig:methods} (a).
  The faint blue curve indicates the steady-state $\textrm{LWP}$ as in Figure~\ref{fig:methods}.
  Fit slopes $\textrm{d}\ln \textrm{LWP}/\textrm{d} \ln N$ are indicated.
  (b) Each data point indicates an adjustment slope obtained as in (a) with error bars for $2\,\textrm{h}\le t \le 12\,$h.
  The red line shows the theoretically expected exponential decay (Equation~\ref{eq:tau}).
  \label{fig:adj_fits}}
\end{figure}

Following the methodology of climatological satellite studies, we derive LWP adjustments $\textrm{d}\ln \textrm{LWP}/\textrm{d} \ln N$ as slopes of linear regression lines through median LWP values in $N$-bins.
To discuss the time-dependence of adjustments, we separately derive LWP-adjustments per time-step.
We illustrate this for $t=2\,$h (magenta data subset in Figure~\ref{fig:methods}, a, and magenta elements in Figure~\ref{fig:adj_fits}, a) and $t=12\,$h (green).
Considering all time steps $2\le t/\textrm{h}\le 12$ shows that the LWP adjustment becomes increasingly negative over time (Figure~\ref{fig:adj_fits}, b). 
This behavior results from the sampling of the $N$-LWP space by our simulations, which evolves over time.
By construction, our dataset initially features an uncorrelated sampling.
This explains the almost horizontal regression line and corresponding vanishing adjustment observed at $t=2\,h$.
An initial co-variability of $N$ and LWP values would have imprinted an initial correlation and and corresponding adjustment value between $N$ and LWP.

As our simulations collectively evolve further from the initial state, they approach the steady-state LWP line (blue curve in Figures~\ref{fig:methods},~\ref{fig:adj_fits}a) and the sampling of the $N$-LWP space features an increasingly negative correlation.
Had we run our simulations for longer than 12\,h, all ensemble members would eventually have reached their steady-state LWP.
This means that for $t\rightarrow \infty$ only the steady-state line is sampled and the LWP adjustment is quantified by the slope of this line.
As the slope of the steady-state LWP line reflects the $N$-dependence of entrainment \cite{HoffmannGlassmeierFeingold19},
the LWP-adjustment at $t \rightarrow \infty$, $\textrm{d} \ln \textrm{LWP}_\infty/\textrm{d} \ln N$, is a direct quantification of $N$-- or more generally aerosol-- effects on cloud processes.

For non-precipitating Sc, we obtain $\textrm{d}\ln\textrm{LWP}_\infty/ \textrm{d} \ln N = -0.64$ (Figure~\ref{fig:adj_fits}, a; for uncertainty quantification see Table~\ref{tab:uncertainty}).
This value constitutes a lower bound;
a more negative adjustment value would require a stronger $N$-dependent entrainment and therefore drier above-cloud conditions than prescribed for our simulations.
This is not realistic since our simulations feature very dry conditions already (Table~\ref{tab:sim_parameters}, Figure~\ref{fig:qt}).
Figure~\ref{fig:lit} also supports $-0.64 \le \textrm{d}\ln \textrm{LWP} / \textrm{d}\ln N$ as a lower bound on previous estimates from the literature.
We contrast this value with the positive value of the precipitation-dominated branch, for which we determine a slope of $0.21$ (Table~\ref{tab:uncertainty}) that lies well within the reported range (Figure~\ref{fig:lit}).

The equilibration of adjustments to the steady-state value is the collective result of the equilibration of individual systems.
This allows us to derive that the observed time-dependence of LWP adjustments is accurately described as an exponential decay towards $\textrm{d}\ln\textrm{LWP}_\infty/\textrm{d} \ln N$ (Figure~\ref{fig:adj_fits}, b),
\begin{eqnarray}
  \nonumber
  \textrm{adj}(\Delta t) &=& \frac{\textrm{d}\ln \textrm{LWP}_\infty}{\textrm{d}\ln N}\left[1-\exp\left(-\frac{\Delta t}{\tau_\textrm{adj}}\right)\right], \\
  \tau_\textrm{adj} &\approx& \tau \left(1 - 1.6\,\frac{\textrm{d}\ln\textrm{LWP}_\infty}{\textrm{d} \ln N}\right) = 2.0\,\tau = 20\,\textrm{h},
  \label{eq:tau}
\end{eqnarray}
with an adjustment equilibration timescale $\tau_\textrm{adj}$
that scales with the equilibration timescale of an individual system, $\tau$ (Figure~\ref{fig:ts}), and with adjustment strength (\emph{Supplementary Information}).
The time-dependence of LWP adjustments on a timescale of almost a day is in stark contrast to the radiative effect of an increased cloud droplet number, which takes full effect in $5-10\,$minutes.

In summary, the extent and interpretation of LWP adjustments in a Sc field depends on the proximity of the system's LWP to its steady-state LWP.
Adjustments based on sampling transient LWP, far from steady state, reflect $N$-LWP co-variability (or the absence thereof in our case) that is externally prescribed on the system--- i.e. a mere \emph{association};
LWP adjustments diagnosed from steady systems reflect aerosol-dependent cloud processes --- i.e. a \emph{causal} relationship;
intermediate degrees of proximity result in a mixture of both.

\subsubsection*{Insufficient time for evolution of ship tracks towards steady state}

\begin{figure}
  \centering
  \begin{overpic}[height=0.39\textwidth]{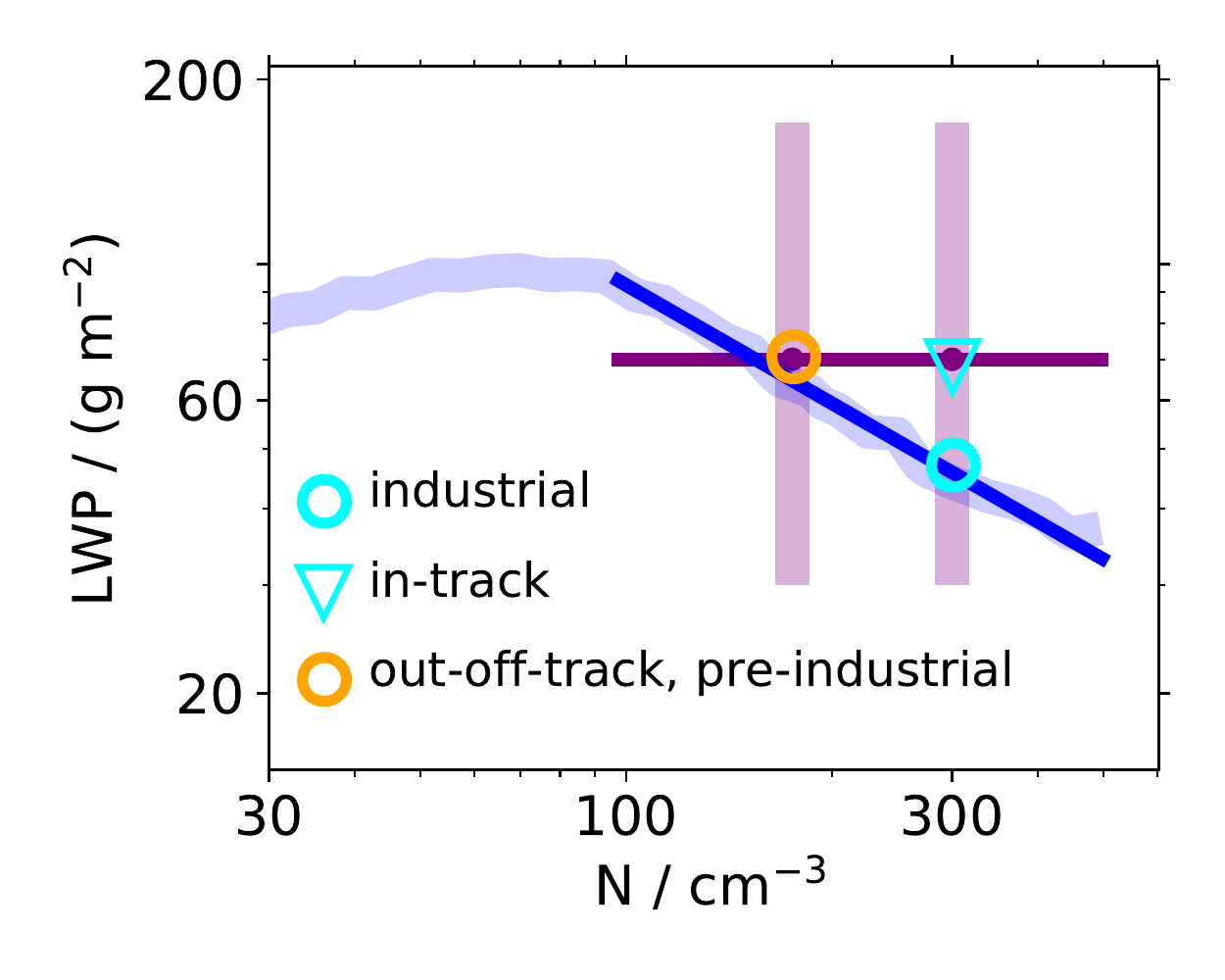}
    \put(69, 52){$\stackrel{(1)}{\rightarrow}$}
    \put(68, 60){$\stackrel{(2a)}{\leftarrow}$}
    \put(81, 44){$\stackrel{}{\downarrow}$}
    \put(85, 43){$\stackrel{(2b)}{}$}
  \end{overpic}
  \caption{Conceptual illustration of LWP adjustments as derived from ship tracks in comparison to climatological satellite studies in analogy to Figure~\ref{fig:adj_fits} (a).
  (Pre-)industrial climatological conditions are strongly-adjusted and represented by circles close to the steady-state (blue line).
  LWPs within ship tracks are weakly adjusted (triangle symbol) such that ship-track studies are based on comparing almost identical LWP distributions (purple; illustration only, no actual data), which implies vanishing LWP adjustments (purple regression line). 
  Labeled arrows correspond to Figure~\ref{fig:cartoon}.
  \label{fig:adj_ship}}
\end{figure}

The degree of proximity of an ensemble, or sampling, of Sc systems to its steady-state LWP adjustment can be estimated by comparing the duration of its evolution under an aerosol perturbation, $\Delta t$, to the characteristic adjustment equilibration timescale, $\tau_\textrm{adj}=20\,$h (Equation~\ref{eq:tau}).
From a Lagrangian perspective, a Sc system is exposed to an aerosol background throughout its lifetime.
Typical Sc trajectories in the subtropics persist on timescales of days, $\Delta t_\textrm{clim} > 48\,$h, before they transition into the shallow cumulus regime due to advection towards higher sea-surface temperatures \cite{Sandu_2011}.
Since $\Delta t_\textrm{clim} \gg \tau_\textrm{adj}$, the climatological sampling of Sc is dominated by strongly equilibrated LWPs.
While not necessarily composed of steady-state LWPs,
we can assume that the LWP climatology of non-precipitating Sc is better characterized as a sampling of steady-state LWPs, than as one of highly transient LWP.
Steady-state values as a feasible approximation for Sc properties are in line with previous theoretical studies \cite{Stevens_2006,BrethertonUchidaBlossey10}.
A significant probability of Sc being observed close to their steady state is also consistent with relatively narrow climatological distributions of Sc LWPs \cite{GryspeerdtGorenSourdeval19, Wood_2012} as transient LWPs are expected to scatter (magenta data in Figure~\ref{fig:methods}, a).
A tendency to rapidly re-equilibrate towards a steady state after a perturbation can furthermore explain the observation of resilient, or buffered, cloud behavior \cite{GlassmeierLohmann18, Wood_2012, StevensFeingold09, ZhuBrethertonKoehler05}.

Sc decks being strongly adjusted to the aerosol background in which they evolve has implications for constraining the anthropogenic radiative forcing;
LWP adjustments need to compare Sc that are strongly adjusted to an aerosol background typical of an industrial-era aerosol climatology (Figure~\ref{fig:adj_ship}, cyan circle at higher $N$) to Sc decks that are strongly adjusted to a pre-industrial aerosol background (orange circle at lower $N$).
Climatological satellite studies are suitable for this quantification because they predominantly sample strongly-adjusted steady-state LWP.
As discussed in the previous section, this specifically means that such studies capture cloud processes, and are only weakly confounded by externally-induced $N$-LWP co-variability.

Ship-track data are obtained throughout the life of the track, with fresh tracks more likely to be sampled due to their better visibility.
With a typical lifetime for ship tracks of $6-7\,$h \cite{ChristensenSuzukiZambri14, DurkeeChartierBrown99}, this corresponds to an average evolution time until sampling of $\Delta t_\textrm{ship} \approx 3$\,h.
As the characteristic equilibration time exceeds the typical evolution time at sampling, $\Delta t_\textrm{ship} \ll \tau_\textrm{adj}$, we conclude that LWPs sampled from ship tracks are not representative of the aerosol-cloud interaction processes, specifically entrainment, that manifest as a Sc system approaches a steady-state LWP.
Instead, their sampling of transient LWPs carries a strong imprint of their specific initial conditions.
To characterize these conditions, we describe ship-track studies as a sampling within two different $N$-bins, one representing out-of-track, and the other in-track conditions (Figure~\ref{fig:adj_ship}).
As LWP adjustments are not instantaneous, the LWP distributions within these two bins are identical when the ship exhaust first makes contact with the cloud.
As for the idealized initial conditions in our dataset, this corresponds to an initial adjustment of zero (Figure~\ref{fig:adj_ship}, purple regression line).
After the perturbation, the in-track distribution evolves to an asymptotic LWP value that is different from that of the out-of-track LWP.
Due to the short duration of this evolution until sampling, adjustment values diagnosed from ship tracks remain small.
Indeed, evolution according to Equation~\ref{eq:tau} corresponds to an adjustment value of
\begin{equation}
  \textrm{adj}(\Delta t_\textrm{ship}=3\,\textrm{h})=-0.1,
  \label{eq:ship_adj}
\end{equation}
which matches reported values ranging from $-0.2$ to $0.0$ (Table~\ref{tab:lit}).
When, in contrast, sampling a climatologically polluted situation, adjustments can evolve to more negative values before being sampled and values of $\textrm{adj}(\Delta t_\textrm{clim}=48\,\textrm{h}) \approx -0.6$, close to the asymptotic value of $-0.64$ (Figure~\ref{fig:adj_fits}, a), are obtained.

While ship exhaust may at first glance seem an intriguing proxy for aerosol conditions typical of the industrial-era aerosol climatology,
it does not perturb the pristine background for a sufficiently long time (Figure~\ref{fig:cartoon}).
In other words, typical LWPs in ship tracks are not comparable to LWPs in Sc that experience a higher aerosol background due to an anthropogenic shift of the aerosol climatology (Figure~\ref{fig:adj_ship}, cyan circle vs triangle).
It needs to be stressed that we only dispute the generalization of LWP adjustments derived from ship track studies to estimate the contribution of LWP adjustments to the overall radiative forcing due to anthropogenic aerosol.
Ship tracks do provide reliable estimates of the radiative forcing of the ship tracks themselves.

\subsubsection*{Implications for the effective radiative forcing due to aerosol-cloud interactions}

The discussions in the previous two sections showed that (i) LWP adjustments become more negative as non-precipitating Sc decks evolve under an aerosol perturbation (Figure~\ref{fig:adj_fits}) and (ii) that the LWP in individual ship tracks experiences less time to adjust to the modified aerosol conditions than a Sc deck under climatological high-aerosol conditions (Figure~\ref{fig:adj_ship}).
In combination, these two findings imply that ship-track-derived LWP adjustments are less negative than the LWP adjustment that a Sc deck under climatologically polluted conditions exhibits.
We contend, therefore, that using ship-track derived adjustment values to estimate the radiative forcing of aerosol-cloud interactions means an underestimation of the absolute effect of LWP adjustments on the radiative forcing.
The negative values of LWP adjustments in non-precipitating Sc mean that an increased aerosol load leads to cloud thinning. The associated reduction in short-wave reflectivity implies a warming effect as a result of LWP adjustments that offsets the cooling associated with cloud brightening (Equation~\ref{eq:sensitivity}).
Ship-track studies underestimate this offsetting warming effect of LWP adjustments (Figure~\ref{fig:cartoon}).
With $-0.64 \le \textrm{d}\ln \textrm{LWP} / \textrm{d}\ln N$ as lower bound for LWP adjustments (Figure~\ref{fig:adj_fits}),
this underestimation corresponds to an overestimation of the cooling effect of aerosols on non-precipitating Sc of up to $200\%$ (\emph{Supplementary Information}).
This specifically includes the possibility of an overall warming, rather than cooling, effect of aerosols on non-precipitating Sc (Figure~\ref{fig:lit}).
Since non-precipitating Sc occur more frequently than precipitating Sc \cite{LeonWangLiu08, PossnerEastmanBender19}, this warming effect offsets the cooling effect of positive LWP adjustments in precipitating Sc in the overall climate effect of Sc.

For sufficiently strong aerosol perturbations, the aerosol-induced cloud thinning will lead to the complete dissipation of individual clouds in the Sc deck.
Such a reduction in cloud fraction through aerosol-enhanced entrainment has previously been reported for a cumulus-under-stratocumulus case \cite{XueFeingold2008}.
Assuming that cloud properties in steady state are climatologically representative,
we estimate that cloud fraction starts to fall below CF $\approx 1$ at $N > 800\,\textrm{cm}^{-3}$ (Figure~\ref{fig:emusus}).
Similar cloud fraction adjustments to aerosol perturbations have recently been discussed for the precipitating Sc regime \cite{Goren_2019, Rosenfeld_2019}. 
As for the precipitating case, where cloud fraction adjustments act in the same direction as LWP adjustments (less aerosol leads to more precipitation, less LWP and break-up), aerosol-entrainment-mediated cloud field dissipation enhances the effects of LWP reduction (more aerosol leads to more entrainment, less LWP and reduced cloud fraction).
While cloud field break-up due to aerosol-precipitation interactions is observed for low aerosol conditions, cloud field dissipation due to aerosol-entrainment interactions occurs for high aerosol conditions.
Cloud fraction adjustments appear particularly relevant under high aerosol conditions because they do not saturate like aerosol effects on cloud albedo (pre-factor $1/N$ in Equation~\ref{eq:sensitivity}, \cite{CarslawLeeReddingtion13}).
Aerosol-entrainment induced cloud dissipation therefore implies the possibility of an even more severe underestimation of the warming effect of adjustments by ship-track studies than estimated here.

Our results are consistent with recent satellite estimates of LWP adjustments in Sc \cite{PossnerEastmanBender19, GryspeerdtGorenSourdeval19}.
Our insight that the effects of external co-variability fade as a Sc system evolves towards its internal steady-state refutes $N$-LWP co-variability as the likely explanation for the strongly negative adjustment values reported.
This rebuts a main argument invoked to question the results of these studies.
At the same time, our modeling results show that strongly negative adjustment values are consistent with existing process understanding.
In combination with the limitations of ship-track derived adjustment values discussed above, we therefore conclude that climatological satellite studies should be assigned more weight for estimating LWP adjustments than ship track studies.
Specifically, values of $\textrm{d}\ln\textrm{LWP}/\textrm{d}\ln N= -0.3$ \cite{PossnerEastmanBender19} to $-0.4$ \cite{GryspeerdtGorenSourdeval19} should be considered possible central values rather than lower bounds as in a recent review \cite{Bellouin_2019}.
Our analysis establishes the steady-state adjustment $\textrm{d}\ln\textrm{LWP}_\infty/\textrm{d}\ln N= -0.64$ as a new lower bound for LWP adjustments in non-precipitating Sc.

This recommendation in particular, and our results in general, are moreover consistent with a recent study that derived LWP adjustments from climatological observations of a heavily frequented shipping lane \cite{DiamondDirectorEastman20}.
This setup provides more persistent pollution than an individual ship track, while still suffering from a certain intermittency of pollution as compared to a climatological perturbation.
We estimate an effective lifetime of ship tracks in a shipping lane of $\Delta t_\textrm{lane}=9\,$h (\emph{Supplementary Information}).
As this time is longer than our estimate for individual ship tracks but shorter than Sc lifetime,
it is not surprising that the shipping lane provides a numerical adjustment value that lies in between those derived from single-ship track studies and fully climatological studies (Figure~\ref{fig:lit}).
Our results therefore reconcile and explain the differing LWP adjustments that have recently been reported \cite{PossnerEastmanBender19, GryspeerdtGorenSourdeval19, TollChristensenQuaas19, DiamondDirectorEastman20}.

Satellite remote sensing of thin and broken clouds remains a challenge, with large uncertainties in retrieved values \cite{FuGirolamoLiang19}.
Despite the support for climatological satellite studies that our results provide, it therefore seems desirable to identify alternatives to ship track studies that allow for a direct observation of aerosol effects.
Our analysis shows that suitable natural experiments feature temporally continuous pollution.
Volcanic emission is an example of a continuously polluting natural experiment, which has so far only been described outside of subtropical Sc regions \cite{TollChristensenQuaas19, Malavelle_2017}.
As another natural experiment with long-lasting pollution, outflows of polluted continental air over the ocean are also of interest \cite{Goren_2019}.
Lastly, controlled experiments suggested in the context of assessing the feasibility of marine cloud brightening \cite{Wood_17} could, by design, provide a sufficiently persistent increase in the aerosol background.
At the same time, our results indicate that an intermittent aerosol perturbation, similar to a ship track, may maximize the cooling effect by keeping compensating adjustments small.

In closing, to successfully quantify the cloud-mediated effect of anthropogenic aerosol on the climate system, there is urgent need to quantify the albedo and LWP responses in both precipitating and non-precipitating regions. This will require careful assessment of the  frequency of occurrence and areal coverage of these regions, with attendant consideration of the temporal nature of the LWP responses. Estimates of aerosol-cloud forcing that ignore the non-precipitating regime are likely to significantly overestimate climate cooling.

\appendix
\section*{Supplementary Information}

\setcounter{table}{0}
\renewcommand{\thetable}{S\arabic{table}}
\setcounter{figure}{0}
\renewcommand{\thefigure}{S\arabic{figure}}
\setcounter{equation}{0}
\renewcommand{\theequation}{S\arabic{equation}}

\subsubsection*{Dataset}

\begin{figure}[h!]
\includegraphics[width=\textwidth]{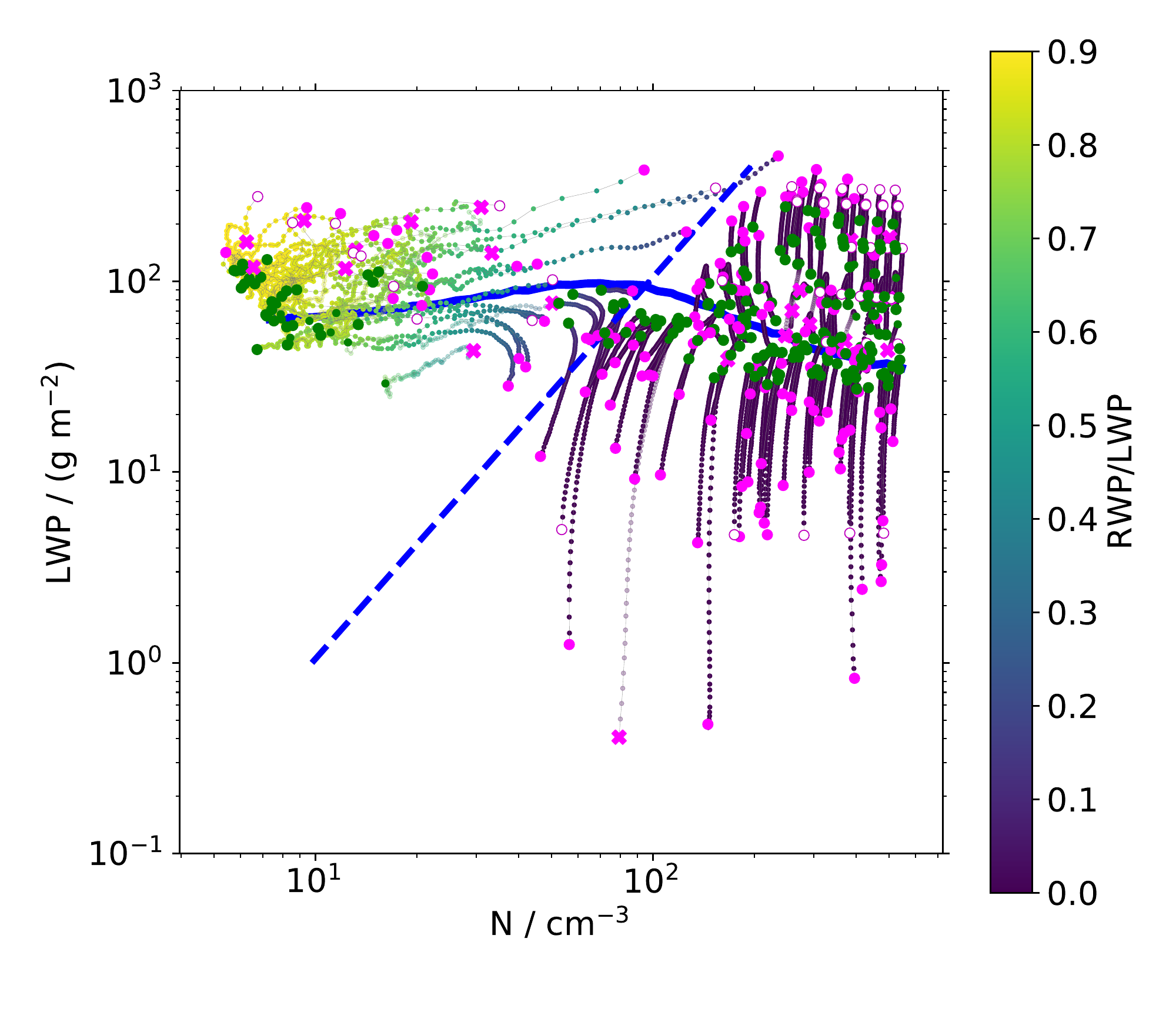}
  \caption{Dataset illustrated similar to Figure~\ref{fig:methods} (a).
  Trajectories in faint coloring, whose start is indicated by a cross rather than a circle, indicate runs that were excluded for this study in comparison to the dataset described in reference \cite{GlassmeierHoffmannJohnson19} due to their above-cloud absolute humidity being an outlier.
  Open magenta-colored circles indicate additional simulations only considered for deriving the flow field $\vec{v}$.
  The coloring of trajectories indicates the fractional contribution of rain water path RWP to total liquid water path LWP.
\label{fig:worms}}
\end{figure}

This study is based on the ensemble of large-eddy simulations (LESs) described in reference \cite{GlassmeierHoffmannJohnson19}.
In comparison to the original dataset, we have excluded outliers in terms of above-cloud humidity, which results in a dataset of 144 LES runs.
External, or large-scale, conditions are the same across the LES ensemble and are summarized in Table~\ref{tab:sim_parameters}.
Variability of LWP within the ensemble is achieved by varying the initial profiles of temperature and moisture;
individual simulations vary in $N$ because they have been initialized with varying aerosol backgrounds (Table~\ref{tab:sim_variables}).
Following reference \cite{FeingoldMcComiskeyYamaguchi16}, we prevent co-variability among the initial conditions by means of a 6D latin-hypercube sampling of the internal factors listed in Table~\ref{tab:sim_variables}.
For the derivation of the flow field, additional simulations have been added to achieve a better coverage of the $N$-LWP space.
The dataset and its variants are illustrated in Figure~\ref{fig:worms}. 

\subsubsection*{Derivation of the flow field $\vec{v}(N, \rm{LWP})$}

\begin{figure}[p]
  \begin{center}
  \begin{overpic}[width=.95\textwidth, trim={0 2cm 0 22cm}, clip]{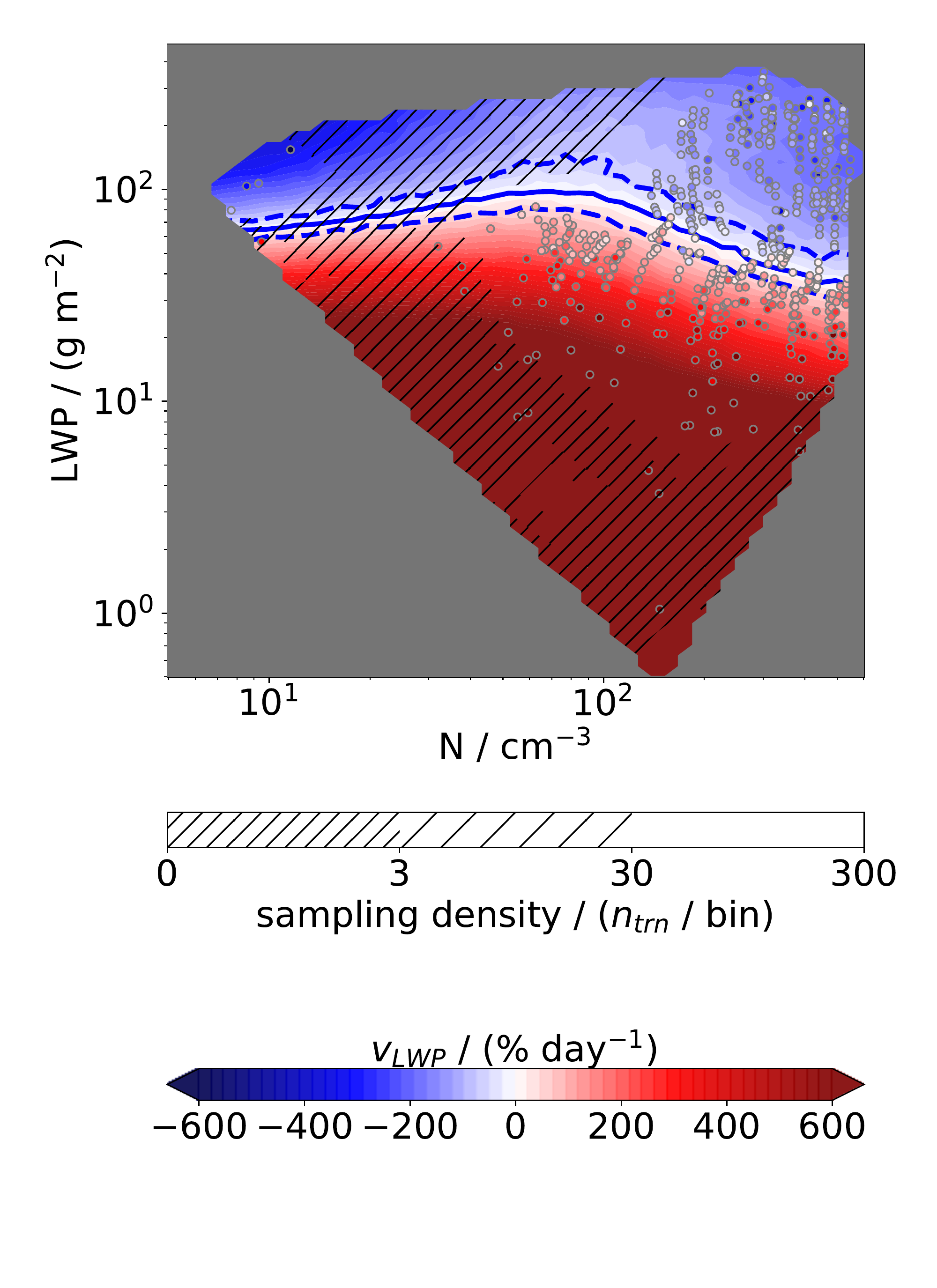}
\put(0,20){a)}
  \end{overpic}
\begin{overpic}[width=.95\textwidth, trim={0 7cm 0 0.7cm}, clip]{sfig2a.pdf}
\put(18.5,48){\includegraphics[scale=0.35]{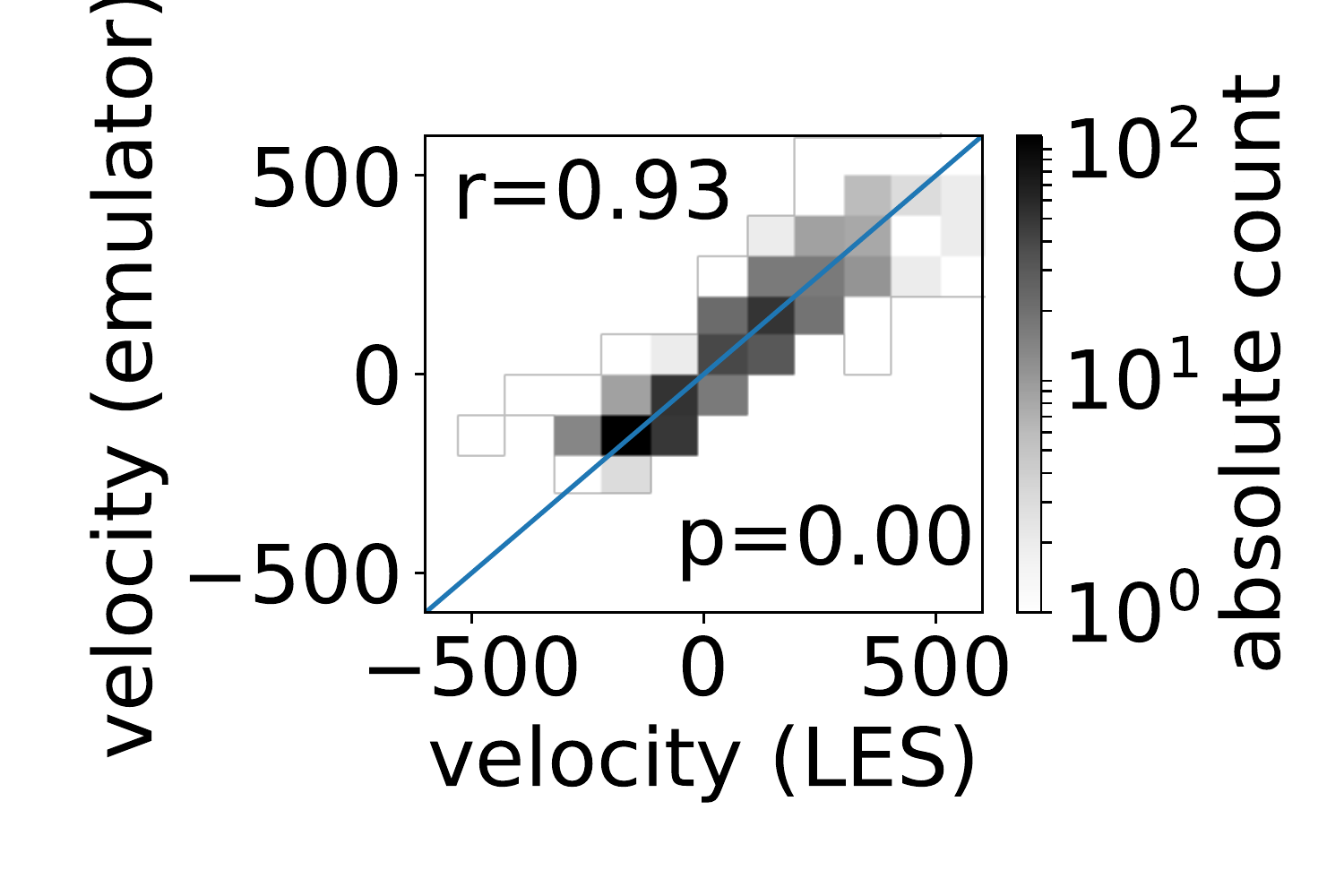}}
\end{overpic}
  \end{center}
  \caption{(See next page for continuation of figure) Emulated surfaces (average surface from an ensemble of emulators) of (a) LWP-component $v_\textrm{LWP}=\textrm{d}\ln \textrm{LWP} / \textrm{d} t$ and (b) $N$-component $v_\textrm{N}=\textrm{d}\ln N / dt$ of the flow field $\vec{v}$ as a function of LWP and droplet number $N$.
  The dark gray area confines the convex hull of data points, within which interpolation is possible. 
The goodness of fit of the emulated surfaces as compared to the validation data set is illustrated by color-filled circles and by the one-to-one scatter plots in the insets, which also indicate a correlation coefficient $r$ and a $p$-value for a linear relationship.
   \label{fig:velocity_components}}
\end{figure}

The flow field $\vec{v} = \left(\textrm{d} \ln N /\textrm{d} \textrm{t}, \textrm{d}\ln\textrm{LWP}/\textrm{d}t\right)^\textrm{T}= \left(v_\textrm{N}, v_\textrm{LWP}\right)^\textrm{T}$ shown in Figure~\ref{fig:methods} (b) is based on separately deriving the components in the $N$-direction, $v_\textrm{N}$, and the LWP-direction, $v_\textrm{LWP}$ (Figure~\ref{fig:velocity_components}).
To derive these component fields, we first extract tendencies $\textrm{d}\ln N/\textrm{d}t$ and $\textrm{d}\ln \textrm{LWP}/\textrm{d}t$ from the data and then interpolate the extracted tendencies by means of \emph{Gaussian-process regression} to obtain emulators for the tendency surfaces.

To derive tendencies, we split each simulated time-series into six intervals of $100\,$min duration, each of which contains 10 consecutive data points at a $10\,$min output frequency. 
For each of these we determine tendencies by fitting trend lines.
We only consider significant trends (p-value $<0.05$) and assign a value of zero otherwise. 
To account for oscillatory behavior that occurs because some of our simulations remain influenced by spin-up processes, we assume that the last $100\,$min-segment of each time-series provides the correct sign of the evolution.
The previous segments are then only considered if they feature the same sign.
With these restrictions, we obtain a dataset of 828 LWP-tendencies and 783 $N$-tendencies.

We process these datasets largely in the same way as described in detail in reference~\cite[Section 3]{GlassmeierHoffmannJohnson19}.
Here we only mention adaptations to the technical parameters mentioned therein.
Instead of a 50\%-50\% split of the dataset into training and validation data, we use a smaller fraction of training data (33\%) to ensure good validation.
To obtain an ensemble of 5 emulated surfaces for $v_\textrm{LWP}$ and $v_\textrm{N}$, we do not restrict the fraction of the training data to be used for individual ensemble members.

\subsubsection*{Quantification and uncertainty of LWP adjustment values in steady state}

As a result of our interpolation technique, we obtain a mean emulator surface $v(N, \textrm{LWP}$), to which individual emulator ensemble members contribute according to their root-mean-square error (RMSE) in predicting the validation data \cite[Equation 3]{GlassmeierHoffmannJohnson19}.
The central value for the LWP adjustment is based on the zero-contour of this mean surface.
For the uncertainty percentiles, we determine LWP adjustments from the zero-contours of a specific sampling of the emulator ensemble.
For this sampling, we take 100 samples of each of the 5 ensemble members and then select a subset of these 500 samples, such that each ensemble member contributes proportionally to its RMSE-based weight, i.e. for the ensemble member with the lowest RMSE, all 100 samples are considered, and for the ensemble member with the highest RMSE, no samples are considered.
Table~\ref{tab:uncertainty} summarizes the uncertainty ranges obtained in this way for the LWP adjustment value.

\subsubsection*{Derivation of the adjustment equilibration timescale}

To derive the adjustment equilibration timescale, we determine the time required by the entire system of LWPs to reach their respective steady states.
We assume that the LWP for any $N$ approaches its steady state $\textrm{LWP}_\infty(N)$ with approximately the same velocity $[\textrm{LWP}_\textrm{ini}-\textrm{LWP}_\infty(N_0)]/\tau$ controlled by the characteristic equilibration timescale $\tau= 9.6\,\textrm{h}$ (Figure~\ref{fig:ts}), where $\textrm{LWP}_\textrm{ini}$ is an initial non-steady-state LWP. 
Hence, the linearized exponential change in LWP yields
\begin{equation}
\textrm{LWP}(t,N) = \textrm{LWP}_\textrm{ini} - \frac{t}{\tau} \left[\textrm{LWP}_\textrm{ini} - \textrm{LWP}_\infty(N_0) \right],
\label{eq:lwp_change}
\end{equation}
where $\textrm{LWP}_\infty(N_0)$ is the steady-state LWP for the smallest $N$ in the non-precipitating regime, $N_0$.
Note that we focus on the LWPs that approach the steady state from larger values since only those require a longer time to reach the steady state for larger $N$.

The steady-state LWP as a function of $N$ is obtained by integrating and linearizing the adjustment $\textrm{d}\ln\textrm{LWP}/\textrm{d} \ln N$,
\begin{equation}
\textrm{LWP}_\infty(N) = \textrm{LWP}_\infty(N_0) \left[1 + \frac{\textrm{d}\ln\textrm{LWP}_\infty}{\textrm{d} \ln N} \ln\left(\frac{N}{N_0}\right)\right],
\label{eq:steady_state_lwp}
\end{equation}
using the same constants of integration as above.
Combining Equations~\ref{eq:steady_state_lwp} and~\ref{eq:lwp_change}, and solving for $t\equiv\tau_\textrm{adj}$, gives the adjustment equilibration timescale necessary to equilibrate the entire system:
\begin{equation}
\tau_\textrm{adj} = \tau \left[1 - \frac{\textrm{d}\ln\textrm{LWP}_\infty}{\textrm{d} \ln N} \ln\left(\frac{N_2}{N_0}\right) \frac{\textrm{LWP}_\infty(N_0)}{\textrm{LWP}_\textrm{ini} - \textrm{LWP}_\infty(N_0)} \right],
\nonumber
\end{equation}
where $N=N_2$ is the largest droplet concentration in the considered non-precipitating regime, resulting in the longest time to equilibrate the LWP.
With $N_0=107\,\textrm{cm}^{-3}$, $N_2=390\,\textrm{cm}^{-3}$ and $\textrm{LWP}_\infty(N_0)=89\,\textrm{g}\,\textrm{m}^{-2}$ in the smallest (index 0) and largest (index 2) $N$-bin,
and $\textrm{d} \ln \textrm{LWP}_\infty / \textrm{d} \ln N=-0.64$,
we obtain best fitting results for $\tau_\textrm{adj}$ when assuming $\textrm{LWP}_\textrm{ini} = 159\,\textrm{g}\,\textrm{m}^{-2}$, which amounts to the 78th percentile of LWPs in the $N_2$-bin.
These numerical values provide the adjustment equilibration timescale stated in Equation~\ref{eq:tau}.

\subsubsection*{Effective evolution time of ship tracks in a shipping lane}

The effective evolution time $\Delta t_\textrm{lane}$ of a ship track in a shipping lane can be estimated based on Equation~\ref{eq:tau},
\begin{equation}
  \textrm{adj}(\Delta t_\textrm{lane}) = -0.24 \Rightarrow \Delta t_\textrm{lane}\approx 9\,\textrm{h},
\end{equation}
where we have used the adjustment value $\textrm{d} \ln \textrm{LWP}/\textrm{d} \ln N = -0.24$ from reference~\cite{DiamondDirectorEastman20}.

\subsubsection*{Cloud-mediated aerosol forcing and cloud radiative effect}

The cloud-mediated aerosol forcing depends on the aerosol sensitivity of the relative cloud radiative effect, $\textrm{rCRE}$, which relates downwelling short-wave radiative fluxes at the surface, $F$, under clear-sky (index $\textrm{clr}$) and all-sky (index $\textrm{all}$) conditions \cite{Xie_2013},
\begin{equation}
\textrm{rCRE}=\frac{F_\textrm{clr}-F_\textrm{all}}{F_\textrm{clr}} \approx \textrm{CF}\cdot A_\textrm{c}\approx A_\textrm{c}
\end{equation}
and amounts to cloud albedo $A_\textrm{c}$ in fully overcast Sc with cloud fraction $\textrm{CF}\approx 1$.
We assume that climatological cloud properties can be approximated by steady-state values.
With a steady-state cloud albedo of $A_\textrm{c}=0.5$ based on Figure~\ref{fig:emusus}, Equation~\ref{eq:sensitivity} for the sensitivity of $A_\textrm{c}$, or rCRE, respectively, results in
\begin{equation}
  S = \frac{1}{N} \left(\frac{1}{12} + \frac{5}{24} \frac{\textrm{d}\ln \textrm{LWP}}{\textrm{d}\ln N}\right).
  \label{eq:rCRE}
\end{equation}

With $\textrm{d}\ln \textrm{LWP}/\textrm{d}\ln N \approx -0.1$ (Equation~\ref{eq:ship_adj}), ship-track studies imply $S_\textrm{ship} \approx 0.06/N>0$ and thus a cooling effect of anthropogenic aerosol via increased cloud brightness at almost constant LWP.
In contrast, the steady-state adjustment value of $\textrm{d}\ln \textrm{LWP}_\infty/\textrm{d}\ln N=-0.64$ derived here as a lower bound results in $S_\textrm{clim}=-0.05/N<0$, which indicates that
aerosol-induced cloud thinning overcompensates the brightening effect at constant LWP.
Ship-track studies thus overestimate the cooling effect of aerosol on Sc by up to $|(-0.05-0.06)/(-0.05)| =220\,\% \approx 200\,\%$.

\newpage
\subsection*{Supplementary figures and tables}

\begin{figure}
  \begin{center}
  \begin{overpic}[width=.95\textwidth, trim={0 2cm 0 22cm}, clip]{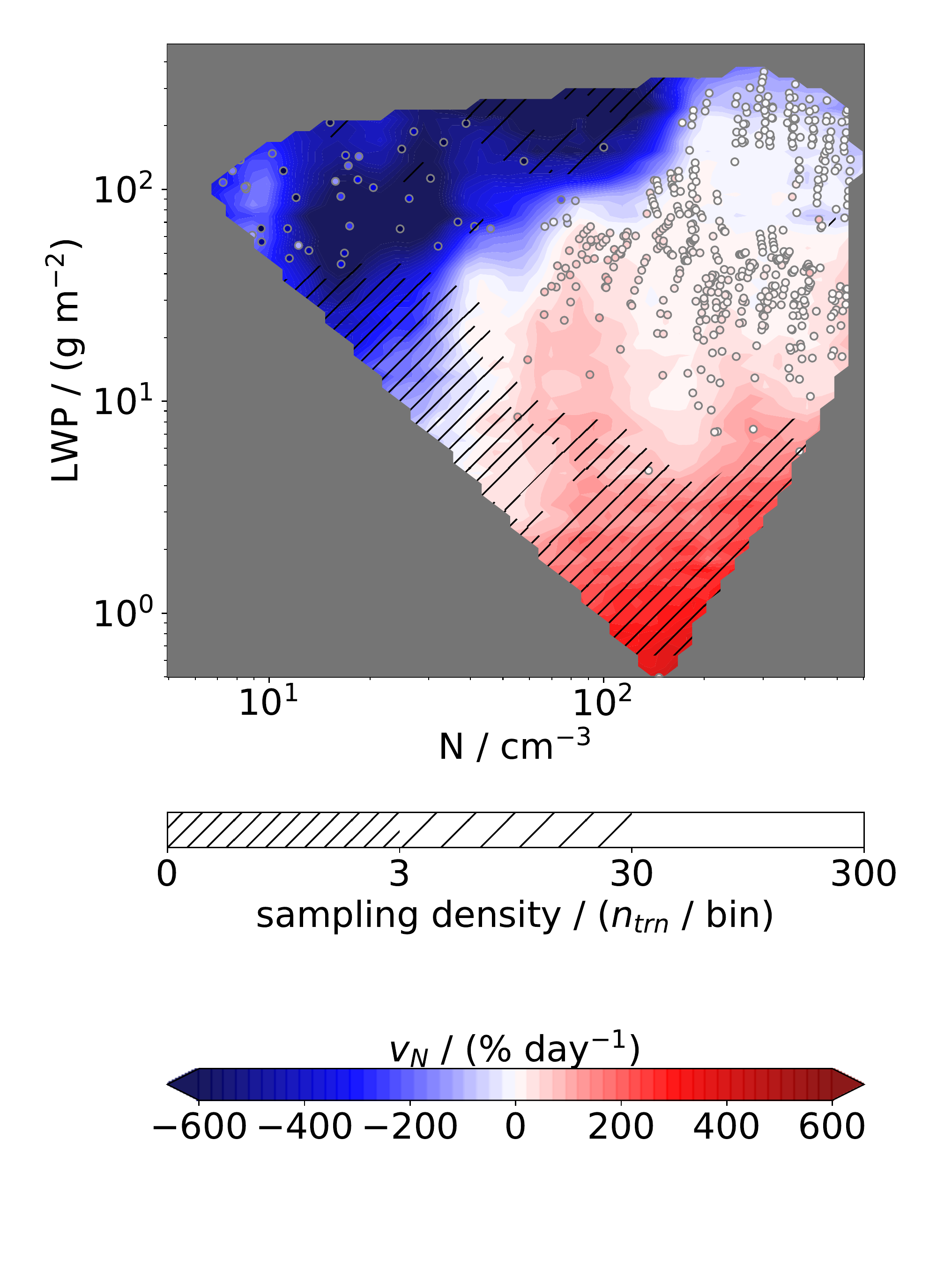}
    \put(0,20){b)}
  \end{overpic}
\begin{overpic}[width=.95\textwidth, trim={0 7cm 0 0.7cm}, clip]{sfig2b.pdf}
\put(18.5,48){\includegraphics[scale=0.35]{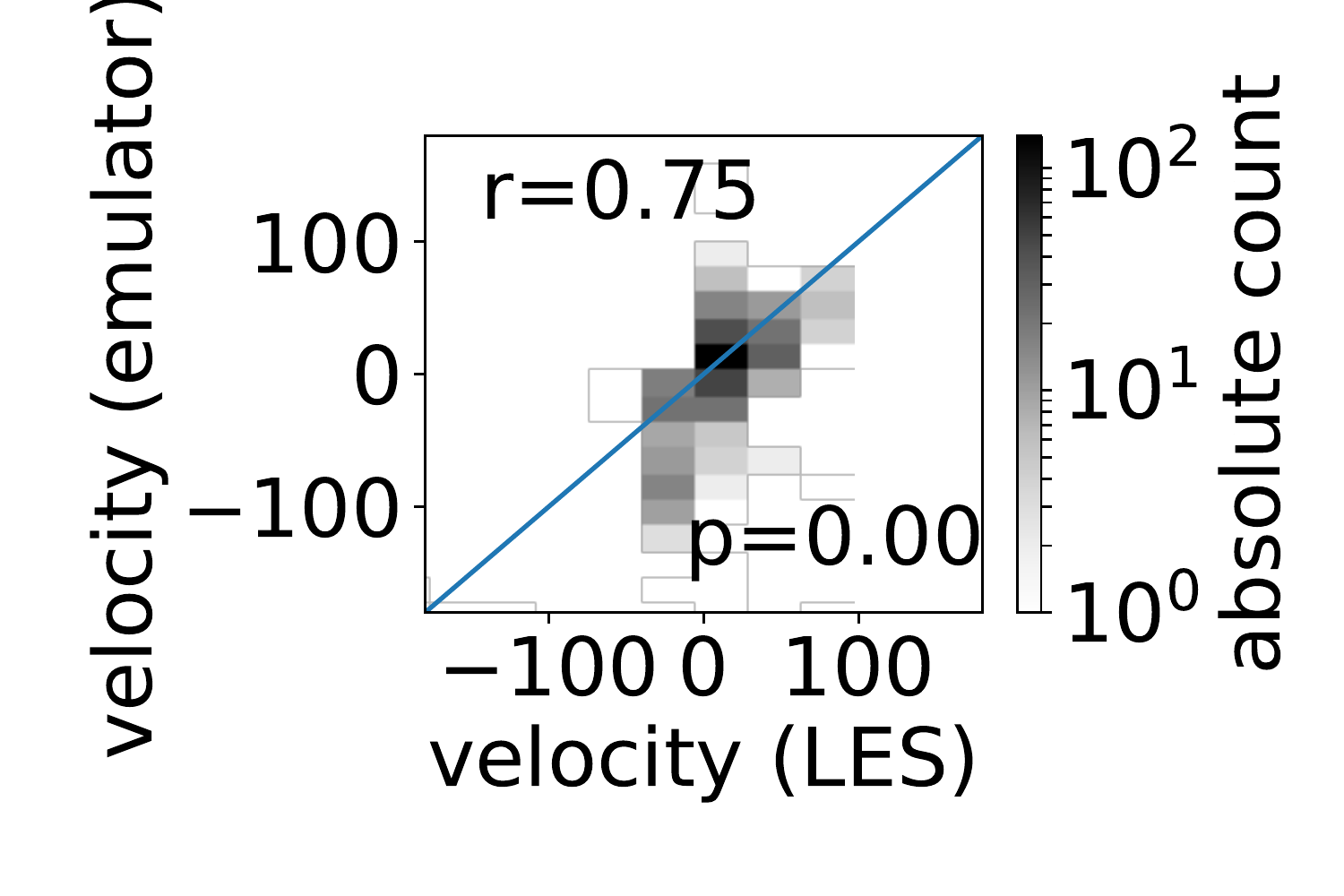}}  
\end{overpic}
  \end{center}

  \phantomcaption{\small(Continuation from previous page)
    Hatching indicates the number of training data points $n_\textrm{trn}$ per bin, for a $10\times12$ binning of the $N$-LWP space.
  As discussed in reference~\cite{GlassmeierHoffmannJohnson19}, insufficient sampling is the largest source of uncertainty.
  Blue contour lines in (a) indicate $v_\textrm{LWP} = 0$, i.e. $\textrm{LWP}=\textrm{LWP}_\infty$, for the (dashed) 25th, (solid) 50th and (dashed) 75th percentile of a RMSE-weighted sampling from the emulator ensemble (see text).
  Blue curves in Figure~\ref{fig:methods} and~\ref{fig:adj_fits} correspond to the median sampling (solid blue contour).
}
\end{figure}

\setcounter{figure}{2}

\begin{figure}
  \begin{center}
    \begin{overpic}[width=.95\textwidth, trim={0 2cm 0 22cm}, clip]{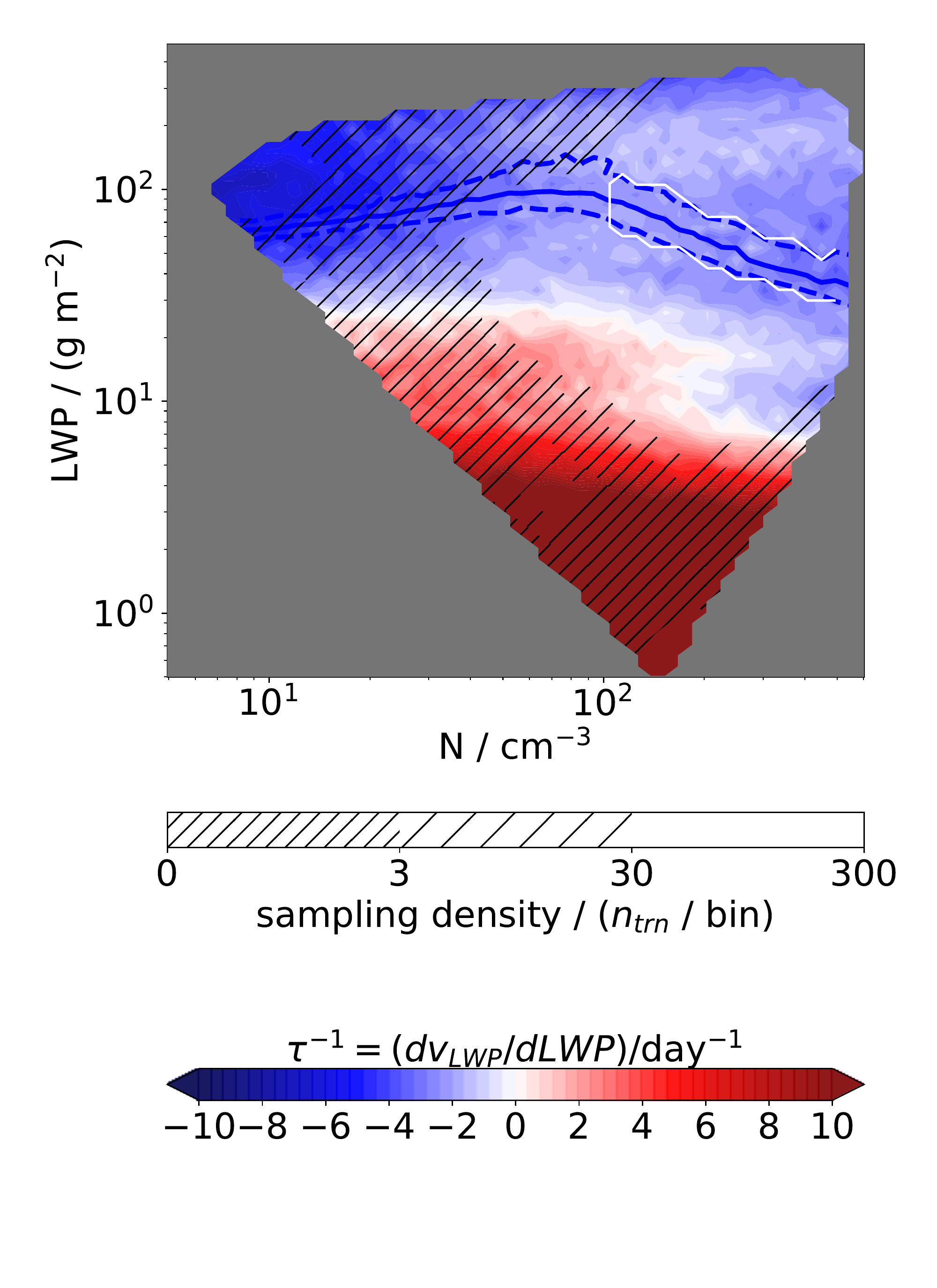}
    \end{overpic}
    \begin{overpic}[width=.95\textwidth, trim={0 7cm 0 0.7cm}, clip]{sfig3.pdf}\end{overpic}
  \end{center}
  \caption{The characteristic equilibration timescale $\tau$ for LWP equilibration to its steady-state value $\textrm{LWP}_\infty$ is determined by $\tau^{-1}=\left |\partial v_\textrm{LWP}/ \partial \textrm{LWP} \right|_{\textrm{LWP}_\infty}$, 
  where the derivative (color contours) is evaluated at $\textrm{LWP}_\infty$ \cite{Strogatz94}.
  An average value of $\partial v_\textrm{LWP}/ \partial \textrm{LWP}=-2.49\,\textrm{day}^{-1}$ within the white contour indicates a characteristic equilibration timescale of $\tau = 9.6\,$h. 
  Sampling density as in Figure~\ref{fig:velocity_components}.
  \label{fig:ts}}
  \vspace{0.43cm}
\end{figure}

\begin{figure}
  \centering
  \includegraphics[width=0.6\textwidth]{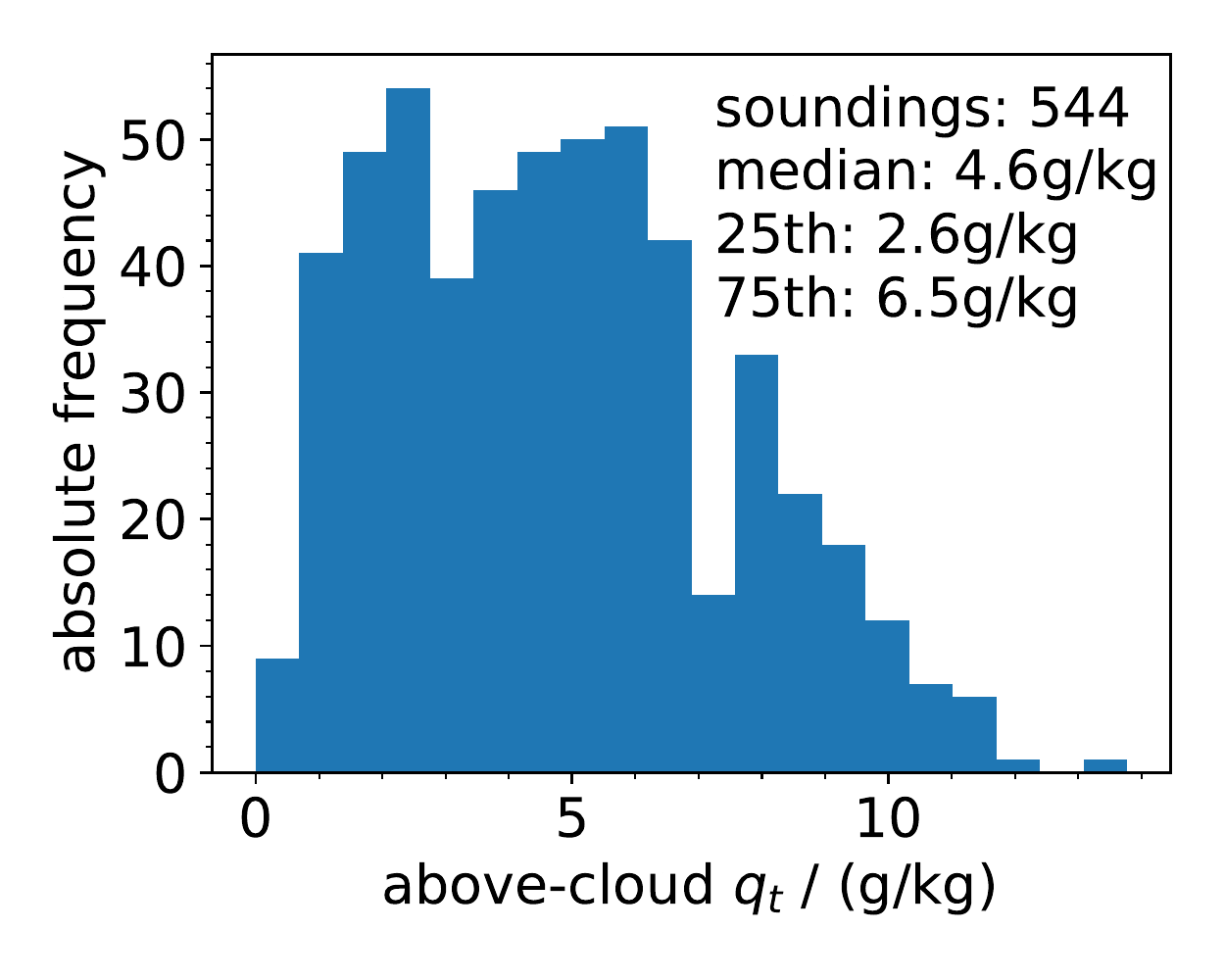}
  \caption{Climatology (1973-2020) of above-cloud mixing ratio at San Clemente Island (Station NSI 72291) in the Sc region off the coast of California.
  Mixing ratios are obtained at a height of $1.05z_\textrm{i}$, where $z_\textrm{i}$ denotes the inversion height as determined from the maximum gradient in the equivalent potential temperature.
  Only soundings that reach saturation in the lowest 2000\,m were considered.
Radiosonde data kindly provided by University of Wyoming (http://www.weather.uwyo.edu/upperair/sounding.html). \label{fig:qt}}
\end{figure}

\begin{figure}
  (a) \hspace{0.45\textwidth} (b)\\[9pt]
  \includegraphics[width=0.49\textwidth]{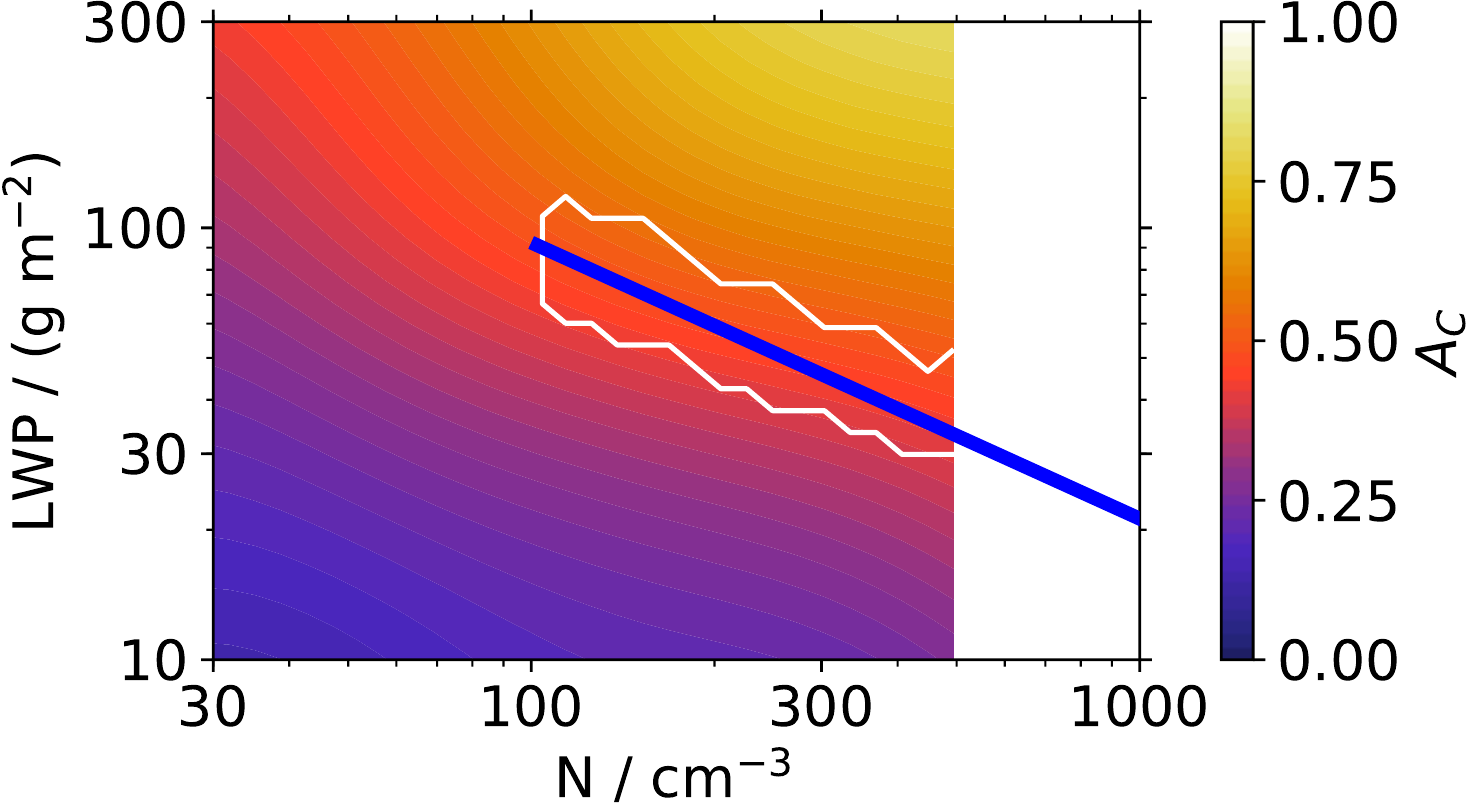}
  \includegraphics[width=0.49\textwidth]{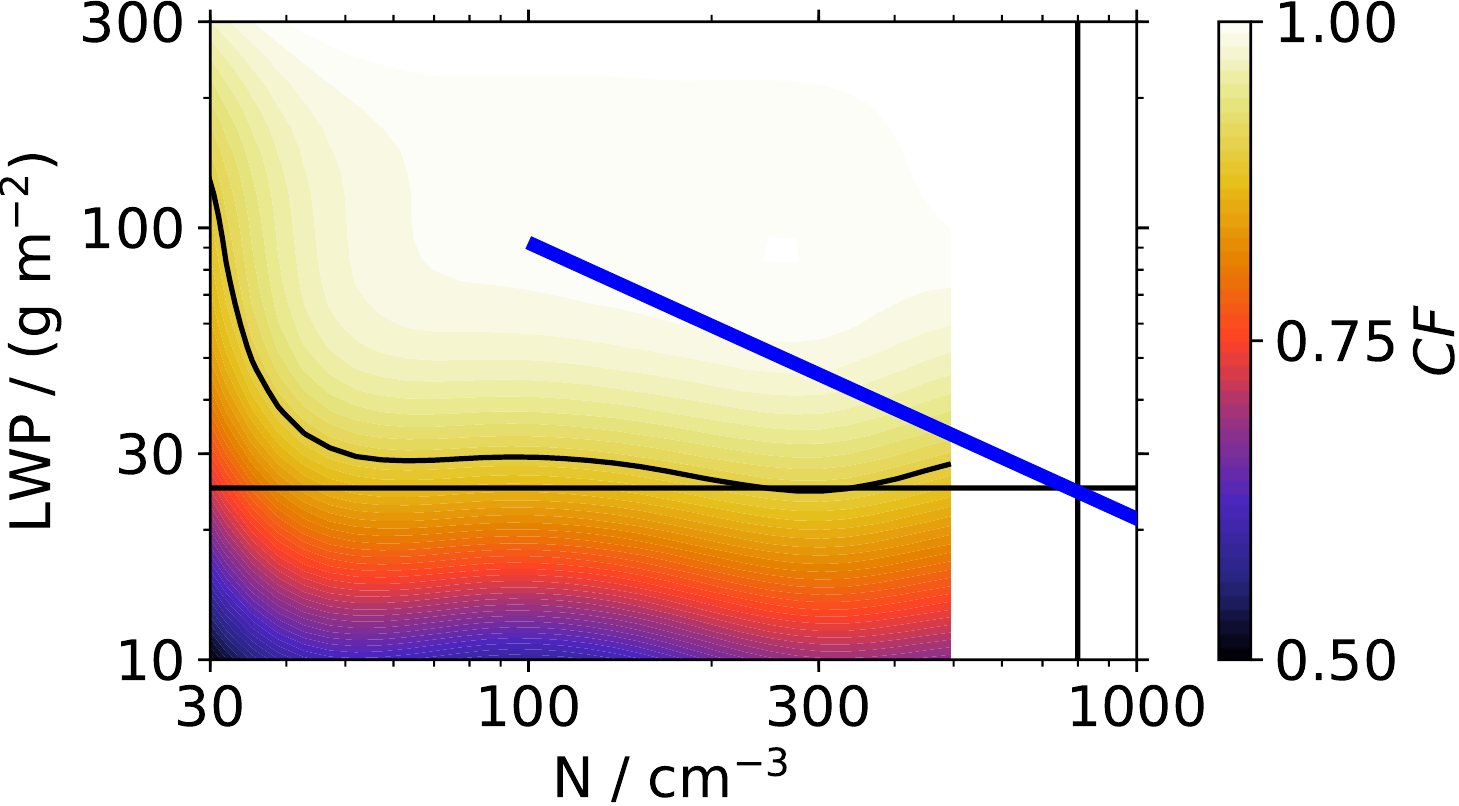}
  \caption{Emulated values for (a) cloud albedo $A_\textrm{c}$ and (b) cloud fraction CF as function of cloud droplet number $N$ and liquid-water path LWP (re-plot of results from reference~\cite{GlassmeierHoffmannJohnson19}).
  Solid blue lines indicate the location of the steady-state LWP as in Figure~\ref{fig:adj_fits}.
  The white contour in (a) corresponds to the blue uncertainty shading in Figure~\ref{fig:methods} (b).
  The average cloud albedo within this white contour amounts to $A_\textrm{c}=0.46$.
  The black contour in (b) indicates $\textrm{CF}=90$\%.
  The horizontal and vertical black lines in (b) are conservative guides to the eye, showing that a cloud fraction reduction $\textrm{CF}<90$\% is expected for $N\approx800\,\textrm{cm}^{-3}$.
  For larger $N$, no fully overcast Sc steady state exists.
  \label{fig:emusus}}
\end{figure}

\begin{landscape}
\begin{table}
  \begin{tabular}{p{0.7cm} | l l p{6cm} l l p{2.8cm}}
    & entr & prcp & reference details & data source & data type & comments\\
    \hline
    this work & -0.64 & 0.21 & Figure~\ref{fig:adj_fits} & LES ensemble & regression & steady-state value\\ 
  \hline
    \cite{Ackermann_2004} & -0.27 & 0.24 & Supplementary Table 1,\newline prcp: ASTEX, FIRE1,\newline entr: DYCOMS-RF1 & LES & difference quotient &  $q_\textrm{t}\approx 1$g/kg \\
  \hline
    \cite{Hill_2009} & -0.05 & - & Table 3, HOM\_FULL & LES & LWP, $N$ (lists) & $q_\textrm{t}>5$g/kg\\
  \hline
    \cite{LeePennerSaleeby09} & 0.11 & - & Figures 4 and 8, MID-WET & LES & LWP, $N$ (pairs) & $q_\textrm{t}=5.5$g/kg\\
  \hline
    \cite{Wang_2011} & -0.12 & 0.24 & Table 2 & LES & LWP, $N$ (pairs) &  $q_\textrm{t}\approx3$\,g/kg \\
  \hline
    \cite{Michibata_2016} & 0.075 & 0.025 & Figure 4, \newline LTS$>$15 K,\newline prcp threshold: $\textrm{dBZe}<-20$ & satellite &  regression & no $N$-binning\\
  \hline
    \cite{GryspeerdtGorenSourdeval19} & -0.42 & 0.14 & Figure 4 (e) & satellite & regression & - \\
  \hline
    \cite{Rosenfeld_2019} & \multicolumn{2}{c}{0.03}& Table 1, LTS$>18$ & satellite & regression & {no separate fits\newline for entr vs prcp}\\
  \hline
    \cite{PossnerEastmanBender19} & -0.31; -0.15 & 0.14&Table 1, $s_\textrm{LWP}$ (bivariate), min and max value for entr & satellite & regression & - \\
  \hline
    \cite{CoakleyWalsh02} & -0.175 & - & abstract & ship track & difference quotient & - \\
  \hline
    \cite{ChristensenStephens11} & -0.13 & 0.36 & Table 2, entr: closed, prcp: open & ship track & LWP, $\tau$, $r_\textrm{eff}$ (pairs) & - \\
  \hline
    \cite{TollChristensenQuaas19} & -0.021 & - & Extended Data Table 2 & ship track & difference quotient & - \\
  \hline
    \cite{DiamondDirectorEastman20} & -0.24 & - & Section 4 & ship track & difference quotient & shipping lane \\
  \end{tabular}
  \caption{Literature values used in Figure~\ref{fig:lit} and details about their derivation.
  Abbreviations ``entr'' and ``prcp'' refer to the entrainment- and precipitation-dominated regimes of Sc, respectively.
  Data source ``satellite'' refers to climatological satellite studies, while ``ship track'' refers to ship track studies from satellite.
  The ``data type'' column indicates whether adjustment information was derived as difference quotient $\Delta \ln \textrm{LWP}/\textrm{d} \ln N$ from value pairs, obtained as gradient $\textrm{d}\ln \textrm{LWP}/\textrm{d}\ln N$ of linear regression lines through a provided list of data points, or directly provided in one of these two forms.
  In one case, $N$ was inferred from cloud optical thickness $\tau$ and effective radius $r_\textrm{eff}$. 
  \label{tab:lit}}
\end{table}
\end{landscape}
\clearpage

\begin{table}[h]
  \centering
  \begin{tabular}{l | c}
    horizontal wind divergence &  $3.75 \times 10^{-6}\,\textrm{s}^{-1}$\\
    \hline
    sensible heat flux& $16\,\textrm{W}\,\textrm{m}^{-2}$ \\
    \hline
    latent heat flux & $93\,\textrm{W}\,\textrm{m}^{-2}$ \\
    \hline
    aerosol surface flux & $70\,\textrm{cm}^{-2}\,\textrm{s}^{-1}$\\
    \hline
    above-cloud moisture & $0.5[0.2,2.8]\,\textrm{g}\,\textrm{kg}^{-1}$\\
    \hline
    radiation & nocturnal 
\end{tabular}
  \caption{External simulation parameters given by large-scale conditions following reference~\cite{AckermanZantenStevens09}. Values of above-cloud moisture refer to the median and, in brackets, the minimum and maximum value within the distribution.\label{tab:sim_parameters}}
\end{table}

\begin{table}[h]
  \centering
\begin{tabular}{l | c}
  mixed-layer height $h$ (in m) & $[500, 1300]$\\
\hline
  mixed-layer aerosol concentration $N_\textrm{a}$  (in cm$^{-3}$) & $[30, 500]$\\
\hline
  mixed-layer liquid-water potential temperature $\theta_l$ (in K)& $[284,294]$\\
\hline
  mixed-layer moisture $q_\textrm{t}$ (in $\textrm{g}\,\textrm{kg}^{-1}$)& $[6.5, 10.5]$\\
\hline
  inversion of liquid-water potential temperature at $h$ $\Delta \theta_\textrm{l}$ (in K) & $[6, 10]$\\
\hline
  inversion of moisture at $h$ $\Delta q_\textrm{t}$ (in $\textrm{g}\,\textrm{kg}^{-1}$) & $[-10, -6]$\\
\end{tabular}
  \caption{Ranges of values that span the initial conditions of internal variables for the ensemble of LES runs used in this study, following reference~\cite{FeingoldMcComiskeyYamaguchi16} and assuming well-mixed initial profiles.\label{tab:sim_variables}}
\end{table}

\begin{table}[h]
  \centering
  \begin{tabular}{l | r r r r r}
   percentile & 5th & 25th & 50th & 75th & 95th\\
    \hline
    entrainment dominated&-1.15& -0.89& -0.64& -0.55&-0.34
    \\
    drizzle dominated &0.17& 0.22 &0.21 & 0.39& 0.68
  \end{tabular}
  \caption{Uncertainty quantification for LWP adjustment values in steady state. See text for details.
  \label{tab:uncertainty}}
\end{table}

\section*{Acknowledgments}
FG thanks Tom Goren and Anna Possner for helpful discussions about the interpretation of satellite literature.
FG acknowledges support by The Branco Weiss Fellowship -- Society in Science, administered by the ETH Z\"urich, and by a Veni grant of the Dutch Research Council (NWO). 
FH holds a visiting fellowship of the Cooperative Institute for Research in Environmental Sciences (CIRES) at the University of Colorado Boulder, and the NOAA/Earth System Research Laboratory.
Jill S. Johnson and Ken S. Carslaw were supported by the Natural Environment Research Council (NERC) under grant NE/I020059/1 (ACID-PRUF) and the UK-China Research and Innovation Partnership Fund through the Met Office Climate Science for Service Partnership (CSSP) China as part of the Newton Fund.
Ken S. Carslaw is currently a Royal Society Wolfson Research Merit Award holder.
This research was partially supported by the Office of Biological and Environmental Research of the U.S. Department of Energy Atmospheric System Research Program Interagency Agreement {DE-SC0016275} and by an Earth's Radiation Budget grant, NOAA CPO Climate \& CI \#03-01-07-001.
Marat Khairoutdinov graciously provided the SAM model.
The University of Wyoming, Department of Atmospheric Science, is acknowledged for archiving the radiosonde data.


\begin{thebibliography}{52}
\providecommand{\natexlab}[1]{#1}
\providecommand{\url}[1]{\texttt{#1}}
\expandafter\ifx\csname urlstyle\endcsname\relax
  \providecommand{\doi}[1]{doi: #1}\else
  \providecommand{\doi}{doi: \begingroup \urlstyle{rm}\Url}\fi

\bibitem[Stephens et~al.(2012)Stephens, Li, Wild, Clayson, Loeb, Kato,
  L'Ecuyer, Stackhouse, and Lebsock]{Stephens_2012}
G.~L. Stephens, J.~Li, M.~Wild, C.~A. Clayson, N.~Loeb, S.~Kato, T.~L'Ecuyer,
  P.~W. Stackhouse, and T.~Lebsock, M.and~Andrews.
\newblock An update on {Earth}'s energy balance in light of the latest global
  observations.
\newblock \emph{Nature Geosci.}, 5\penalty0 (10):\penalty0 691--696, 2012.
\newblock ISSN 1752-0908.
\newblock \doi{10.1038/ngeo1580}.

\bibitem[L'Ecuyer et~al.(2019)L'Ecuyer, Hang, Matus, and Wang]{L_Ecuyer_2019}
T.~S. L'Ecuyer, Y.~Hang, A.~V. Matus, and Z.~Wang.
\newblock Reassessing the effect of cloud type on {Earth}'s energy balance in
  the age of active spaceborne observations. {Part I}: Top of atmosphere and
  surface.
\newblock \emph{J. Climate}, 32\penalty0 (19):\penalty0 6197--6217, 2019.
\newblock \doi{10.1175/jcli-d-18-0753.1}.

\bibitem[Boucher et~al.(2013)Boucher, Randall, Artaxo, Bretherton, Feingold,
  Forster, Kerminen, Kondo, Liao, Lohmann, Rasch, Satheesh, Sherwood, Stevens,
  and Zhang]{BoucherRandall13}
O.~Boucher, D.~Randall, P.~Artaxo, C.~Bretherton, G.~Feingold, P.~Forster,
  V.-M. Kerminen, Y.~Kondo, H.~Liao, U.~Lohmann, P.~Rasch, S.K. Satheesh,
  S.~Sherwood, B.~Stevens, and X.Y. Zhang.
\newblock Clouds and aerosols.
\newblock In T.~F. Stocker, D.~Qin, G.-K. Plattner, M.~Tignor, S.K. Allen,
  J.~Boschung, A.~Nauels, Y.~Xia, V.~Bex, and P.M. Midgley, editors,
  \emph{Climate Change 2013: The Physical Science Basis. Contribution of
  Working Group I to IPCC AR5}. Cambridge, 2013.
\newblock \doi{10.1017/CBO9781107415324}.

\bibitem[Bellouin et~al.(2019)Bellouin, Quaas, Kinne, Stier, Watson-Parris,
  Boucher, Carslaw, Christensen, Daniau, Dufresne, Feingold, Fiedler, Forster,
  Gettelman, Haywood, Lohmann, Malavelle, Mauritsen, McCoy, Myhre,
  {M\"ul\-men\-st\"adt}, Neubauer, Possner, Rugenstein, Sato, Schulz, Schwartz,
  Sourdeval, Storelvmo, Toll, Winker, and Stevens]{Bellouin_2019}
N.~Bellouin, J.~Quaas, E.~Gryspeerdtand~S. Kinne, P.~Stier, D.~Watson-Parris,
  O.~Boucher, K.~S. Carslaw, M.~Christensen, A.-L. Daniau, J.-L. Dufresne,
  G.~Feingold, S.~Fiedler, P.~Forster, A.~Gettelman, J.~M. Haywood, U.~Lohmann,
  F.~Malavelle, T.~Mauritsen, D.~T. McCoy, G.~Myhre, J.~{M\"ul\-men\-st\"adt},
  D.~Neubauer, A.~Possner, M.~Rugenstein, Y.~Sato, M.~Schulz, S.~E. Schwartz,
  O.~Sourdeval, T.~Storelvmo, V.~Toll, D.~Winker, and B.~Stevens.
\newblock Bounding global aerosol radiative forcing of climate change.
\newblock \emph{Rev. Geophys.}, 58\penalty0 (e2019RG000660), 2019.
\newblock \doi{1029/2019RG000660}.

\bibitem[Twomey(1974)]{Twomey74}
S.~Twomey.
\newblock Pollution and the planetary albedo.
\newblock \emph{Atmos. Environ.}, 8:\penalty0 1251--1256, 1974.
\newblock \doi{10.1016/0004-6981(74)90004-3}.

\bibitem[Albrecht(1989)]{Albrecht89}
B.~A. Albrecht.
\newblock Aerosols, cloud microphysics, and fractional cloudiness.
\newblock \emph{Science}, 245:\penalty0 1227--1230, 1989.
\newblock \doi{10.1126/science.245.4923.1227}.

\bibitem[Wang et~al.(2003)Wang, Wang, and Feingold]{WangWangFeingold03}
S.~Wang, Q.~Wang, and G.~Feingold.
\newblock Turbulence, condensation, and liquid water transport in numerically
  simulated nonprecipitating stratocumulus clouds.
\newblock \emph{J. Atmos. Sci.}, 60:\penalty0 262--278, 2003.

\bibitem[Ackerman et~al.(2004)Ackerman, Kirkpatrick, Stevens, and
  Toon]{Ackermann_2004}
A.~S. Ackerman, M.~P. Kirkpatrick, D.~E. Stevens, and O.~B. Toon.
\newblock The impact of humidity above stratiform clouds on indirect aerosol
  climate forcing.
\newblock \emph{Nature}, 432\penalty0 (7020):\penalty0 1011--4, 2004.
\newblock \doi{10.1038/nature03137}.

\bibitem[Bretherton et~al.(2007)Bretherton, Blossey, and
  Uchida]{BrethertonBlosseyUchida07}
C.~S. Bretherton, P.~N. Blossey, and J.~Uchida.
\newblock Cloud droplet sedimentation, entrainment efficiency, and subtropical
  stratocumulus albedo.
\newblock \emph{Geophys. Res. Lett.}, 34\penalty0 (L03813), 2007.

\bibitem[Small et~al.(2009)Small, Chuang, Feingold, and
  Jiang]{SmallChuangFeingold09}
J.~D. Small, P.~Y. Chuang, G.~Feingold, and H.~Jiang.
\newblock Can aerosol decrease cloud lifetime?
\newblock \emph{Geophys. Res. Lett.}, 36\penalty0 (L16806), 2009.

\bibitem[Hoffmann and Feingold(2019)]{Hoffmann_2019}
F.~Hoffmann and G.~Feingold.
\newblock Entrainment and mixing in stratocumulus: Effects of a new explicit
  subgrid-scale scheme for large-eddy simulations with particle-based
  microphysics.
\newblock \emph{J. Atmos. Sci.}, 76\penalty0 (7):\penalty0 1955--1973, 2019.
\newblock \doi{10.1175/jas-d-18-0318.1}.

\bibitem[Stevens and Feingold(2009)]{StevensFeingold09}
B.~Stevens and G.~Feingold.
\newblock Untangling aerosol effects on clouds and precipitation in a buffered
  system.
\newblock \emph{Nature}, 461:\penalty0 607--613, 2009.
\newblock \doi{10.1038/nature08281}.

\bibitem[{M\"ul\-men\-st\"adt} and Feingold(2018)]{MuelmenstaedtFeingold18}
J.~{M\"ul\-men\-st\"adt} and G.~Feingold.
\newblock The radiative forcing of aerosol-cloud interactions in liquid clouds:
  wrestling and embracing uncertainty.
\newblock \emph{Current Climate Change Reports}, 4\penalty0 (1):\penalty0
  23--40, 2018.
\newblock \doi{doi.org/10.1007/s40641-018-0089-y}.

\bibitem[Christensen and Stephens(2011)]{ChristensenStephens11}
M.~W. Christensen and G.~L. Stephens.
\newblock Microphysical and macrophysical responses of marine stratocumulus
  polluted by underlying ships: evidence of cloud deepening.
\newblock \emph{J. Geophys. Res.}, 116\penalty0 (D03201), 2011.

\bibitem[Gryspeerdt et~al.(2019)Gryspeerdt, Goren, Sourdeval, Quaas,
  {M\"ul\-men\-st\"adt}, Dipu, Unglaub, Gettelman, and
  Christensen]{GryspeerdtGorenSourdeval19}
E.~Gryspeerdt, T.~Goren, O.~Sourdeval, J.~Quaas, J.~{M\"ul\-men\-st\"adt},
  S.~Dipu, C.~Unglaub, A.~Gettelman, and M.~Christensen.
\newblock Constraining the aerosol influence on cloud liquid water path.
\newblock \emph{Atmos. Chem. Phys.}, 19:\penalty0 5331--5347, 2019.

\bibitem[Toll et~al.(2019)Toll, Christensen, Quaas, and
  Bellouin]{TollChristensenQuaas19}
V.~Toll, M.~Christensen, J.~Quaas, and N.~Bellouin.
\newblock Weak average liquid-cloud-water response to anthropogenic aerosols.
\newblock \emph{Nature}, 572:\penalty0 51--, 2019.

\bibitem[Diamond et~al.(2020)Diamond, Director, Eastman, Possner, and
  Wood]{DiamondDirectorEastman20}
M.~S. Diamond, H.~M. Director, R.~Eastman, A.~Possner, and R.~Wood.
\newblock Substantial cloud brightening from shipping in subtropical low
  clouds.
\newblock \emph{AGU Advances}, 1\penalty0 (e2019AV000111), 2020.
\newblock \doi{10.1029/2019AV000111}.

\bibitem[Platnick and Twomey(1994)]{PlatnickTwomey94}
S.~Platnick and S.~Twomey.
\newblock Determining the susceptibility of cloud albedo to changes in droplet
  concentration with the advanced very high resolution radiometer.
\newblock \emph{J. Appl. Meteor.}, 33:\penalty0 334 -- 347, 1994.

\bibitem[Boers and Mitchell(1994)]{Boers_1994}
R.~Boers and R.~M. Mitchell.
\newblock Absorption feedback in stratocumulus clouds influence on cloud top
  albedo.
\newblock \emph{Tellus A: Dynamic Meteorology and Oceanography}, 46\penalty0
  (3):\penalty0 229--241, Jan 1994.
\newblock \doi{10.3402/tellusa.v46i3.15476}.

\bibitem[{Coakley Jr.} and Walsh(2002)]{CoakleyWalsh02}
J.~A. {Coakley Jr.} and C.~D. Walsh.
\newblock Limits to aerosol indirect radiative effect derived from observations
  of ship tracks.
\newblock \emph{J. Atmos. Sci.}, 59:\penalty0 668--680, 2002.

\bibitem[Hill et~al.(2009)Hill, Feingold, and Jiang]{Hill_2009}
A.~A. Hill, G.~Feingold, and H.~Jiang.
\newblock The influence of entrainment and mixing assumption on aerosol--cloud
  interactions in marine stratocumulus.
\newblock \emph{J. Atmos. Sci.}, 66\penalty0 (5):\penalty0 1450--1464, 2009.
\newblock ISSN 1520-0469.
\newblock \doi{10.1175/2008jas2909.1}.

\bibitem[Lee et~al.(2009)Lee, Penner, and Saleeby]{LeePennerSaleeby09}
S.~S. Lee, J.~E. Penner, and S.~M. Saleeby.
\newblock Aerosol effects on liquid-water path of thin stratocumulus clouds.
\newblock \emph{J. Geophys. Res.}, 114\penalty0 (D07204), 2009.

\bibitem[Wang et~al.(2011)Wang, Rasch, and Feingold]{Wang_2011}
H.~Wang, P.~J. Rasch, and G.~Feingold.
\newblock Manipulating marine stratocumulus cloud amount and albedo: a
  process-modelling study of aerosol-cloud-precipitation interactions in
  response to injection of cloud condensation nuclei.
\newblock \emph{Atmos. Chem. Phys.}, 11\penalty0 (9):\penalty0 4237--4249,
  2011.
\newblock ISSN 1680-7324.
\newblock \doi{10.5194/acp-11-4237-2011}.

\bibitem[Michibata et~al.(2016)Michibata, Suzuki, Sato, and
  Takemura]{Michibata_2016}
T.~Michibata, K.~Suzuki, Y.~Sato, and T.~Takemura.
\newblock The source of discrepancies in aerosol--cloud--precipitation
  interactions between gcm and a-train retrievals.
\newblock \emph{Atmos. Chem. Phys.}, 16\penalty0 (23):\penalty0 15413--15424,
  2016.
\newblock \doi{10.5194/acp-16-15413-2016}.

\bibitem[Rosenfeld et~al.(2019)Rosenfeld, Zhu, Wang, Zheng, Goren, and
  Yu]{Rosenfeld_2019}
D.~Rosenfeld, Y.~Zhu, M.~Wang, Y.~Zheng, T.~Goren, and S.~Yu.
\newblock Aerosol-driven droplet concentrations dominate coverage and water of
  oceanic low-level clouds.
\newblock \emph{Science}, 363\penalty0 (6427):\penalty0 eaav0566, 2019.
\newblock ISSN 1095-9203.
\newblock \doi{10.1126/science.aav0566}.

\bibitem[Possner et~al.(2019)Possner, Eastman, Bender, and
  Glassmeier]{PossnerEastmanBender19}
A.~Possner, R.~Eastman, F.~Bender, and F.~Glassmeier.
\newblock Deconvolution of boundary layer depth and aerosol constraints on
  cloud water path in subtropical stratocumuli.
\newblock \emph{Atmos. Chem. Phys. Discuss.}, 2019.
\newblock \doi{10.5194/acp-2019-833}.

\bibitem[Leon et~al.(2008)Leon, Wang, and Liu]{LeonWangLiu08}
D.~C. Leon, Z.~Wang, and D.~Liu.
\newblock Climatology of drizzle in marine boundary layer clouds based on 1
  year of data from cloudsat and cloud-aerosol lidar and infrared pathfinder
  satellite observations ({CALIPSO}).
\newblock \emph{J. Geophys. Res.}, 113\penalty0 (D00A14), 2008.
\newblock \doi{10.1029/2008JD009835}.

\bibitem[Xue et~al.(2008)Xue, Feingold, and Stevens]{XueFeingold2008}
H.~Xue, G.~Feingold, and B.~Stevens.
\newblock Aerosol effects on clouds, precipitation, and the organization of
  shallow cumulus convection.
\newblock \emph{J. Atmos. Sci.}, 65\penalty0 (2):\penalty0 392--406, 2008.
\newblock \doi{10.1175/2007jas2428.1}.

\bibitem[Chen et~al.(2014)Chen, Christensen, Stephens, and
  Seinfeld]{ChenChristensenStephens14}
Yi-Chun Chen, M.~W. Christensen, G.~L. Stephens, and J.~H. Seinfeld.
\newblock Satellite-based estimate of global aerosol-cloud radiative forcing by
  marine warm clouds.
\newblock \emph{Nature Geosci.}, 7:\penalty0 643--646, 2014.

\bibitem[Bender et~al.(2019)Bender, Frey, McCoy, Grosvenor, and
  Mohrmann]{BenderFreyMcCoy19}
F.~A.-M. Bender, L.~Frey, D.~T. McCoy, D.~P. Grosvenor, and J.~K. Mohrmann.
\newblock Assessment of aerosol-cloud-radiation correlations in satellite
  observations, climate models and reanalysis.
\newblock \emph{Clim. Dyn.}, 52:\penalty0 4371--4392, 2019.

\bibitem[Brenguier et~al.(2003)Brenguier, Pawlowska, and
  {Sch{\"u}ller}]{BrenguierPawlowskaSchuller03}
J.-L. Brenguier, H.~Pawlowska, and L.~{Sch{\"u}ller}.
\newblock Cloud microphysical and radiative properties for parameterization and
  satellite monitoring of the indirect effect of aerosol on climate.
\newblock \emph{J. Geophys. Res.}, 2003.

\bibitem[Glassmeier et~al.(2019)Glassmeier, Hoffmann, Johnson, Yamaguchi,
  Carslaw, and Feingold]{GlassmeierHoffmannJohnson19}
F.~Glassmeier, F.~Hoffmann, J.~S. Johnson, T.~Yamaguchi, K.~S. Carslaw, and
  G.~Feingold.
\newblock An emulator approach to statocumulus susceptibility.
\newblock \emph{Atmos. Chem. Phys.}, 19:\penalty0 10191--10203, 2019.
\newblock \doi{10.5194/acp-19-10191-2019}.

\bibitem[Feingold et~al.(2016)Feingold, McComiskey, Yamaguchi, Johnson,
  Carslaw, and Schmidt]{FeingoldMcComiskeyYamaguchi16}
G.~Feingold, A.~McComiskey, T.~Yamaguchi, J.~S Johnson, K.~S. Carslaw, and
  K.~S. Schmidt.
\newblock New approaches to quantifying aerosol influence on the cloud
  radiative effect.
\newblock \emph{Proc. Natl. Acad. Sci. USA}, 113\penalty0 (21):\penalty0
  5812--5819, 2016.
\newblock \doi{10.1073/pnas.1514035112}.

\bibitem[Hoffmann et~al.(2020)Hoffmann, Glassmeier, Yamaguchi, and
  Feingold]{HoffmannGlassmeierFeingold19}
F.~Hoffmann, F.~Glassmeier, T.~Yamaguchi, and G.~Feingold.
\newblock Liquid water path steady states in stratocumulus: Insights from
  process-level emulation and mixed-layer theory.
\newblock \emph{J. Atmos. Sci.}, 2020.
\newblock \doi{https://doi.org/10.1175/JAS-D-19-0241.1}.

\bibitem[Rosenfeld and Gutman(1994)]{RosenfeldGutman94}
D.~Rosenfeld and G.~Gutman.
\newblock Retrieving microphysical properties near the tops of potential rain
  clouds by multispectral analysis of {AVHRR} data.
\newblock \emph{Atmos. Res.}, 34:\penalty0 259--283, 1994.

\bibitem[Strogatz(1994)]{Strogatz94}
S.~H. Strogatz.
\newblock \emph{Nonlinear dynamics and chaos}.
\newblock Addison-Wesley, 1994.

\bibitem[Schubert et~al.(1979)Schubert, Wakefield, Steiner, and
  Cox]{SchubertWakefieldSteiner79}
W.~H. Schubert, J.~S. Wakefield, W.~J. Steiner, and S.~K. Cox.
\newblock Marine stratocumulus convection. {Part II:} horizontally
  inhomogeneous solutions.
\newblock \emph{J. Atmos. Sci.}, 36:\penalty0 1308--1324, 1979.

\bibitem[Sandu and Stevens(2011)]{Sandu_2011}
I.~Sandu and B.~Stevens.
\newblock On the factors modulating the stratocumulus to cumulus transitions.
\newblock \emph{J. Atmos. Sci.}, 68\penalty0 (9):\penalty0 1865--1881, 2011.
\newblock \doi{10.1175/2011jas3614.1}.

\bibitem[Stevens(2006)]{Stevens_2006}
B.~Stevens.
\newblock Bulk boundary-layer concepts for simplified models of tropical
  dynamics.
\newblock \emph{Theor. Comput. Fluid. Dyn.}, 20\penalty0 (5-6):\penalty0
  279--304, 2006.
\newblock \doi{10.1007/s00162-006-0032-z}.

\bibitem[Bretherton et~al.(2010)Bretherton, Uchida, and
  Blossey]{BrethertonUchidaBlossey10}
C.~S. Bretherton, J.~Uchida, and T.~N. Blossey.
\newblock Slow manifolds and multiple equilibria in stratocumulus-capped
  boundary layers.
\newblock \emph{J. Adv. Model. Earth Syst.}, 2\penalty0 (14):\penalty0 20,
  2010.
\newblock \doi{10.3894/JAMES.2010.2.14}.

\bibitem[Wood(2012)]{Wood_2012}
R.~Wood.
\newblock Stratocumulus clouds.
\newblock \emph{Mon. Wea. Rev.}, 140\penalty0 (8):\penalty0 2373--2423, 2012.
\newblock \doi{10.1175/mwr-d-11-00121.1}.

\bibitem[Glassmeier and Lohmann(2018)]{GlassmeierLohmann18}
F.~Glassmeier and U.~Lohmann.
\newblock Precipitation susceptibility and aerosol buffering of warm and
  mixed-phase orographic clouds in idealized simulations.
\newblock \emph{J. Atmos. Sci.}, 75\penalty0 (4):\penalty0 1173--1194, 2018.
\newblock \doi{10.1175/JAS-D-17-0254.1}.

\bibitem[Zhu et~al.(2005)Zhu, Bretherton, Koehler, Cheng, Chlond, Geng, Austin,
  Golaz, Lenderink, Lock, and Stevens]{ZhuBrethertonKoehler05}
P.~Zhu, C.~S. Bretherton, M.~Koehler, A.~Cheng, A.~Chlond, Q.~Geng, P.~Austin,
  J.-C. Golaz, G.~Lenderink, A.~Lock, and B.~Stevens.
\newblock Intercomparison and interpretation of single-column model simulations
  of a nocturnal stratocumulus-tropped marine boundary layer.
\newblock \emph{Mon. Wea. Rev.}, 133:\penalty0 2741--2758, 2005.

\bibitem[Christensen et~al.(2014)Christensen, Suzuki, Zambri, and
  Stephens]{ChristensenSuzukiZambri14}
M.~W. Christensen, K.~Suzuki, B.~Zambri, and G.~L. Stephens.
\newblock Ship track observations of a reduced shortwave aerosol indirect
  effect in mixed-phase clouds.
\newblock \emph{Geophys. Res. Lett.}, 4\penalty0 (6970--6977), 2014.

\bibitem[Durkee et~al.(2000)Durkee, Chartier, Brown, Trehubenko, Rogerson,
  Skupniewicz, and Nielsen]{DurkeeChartierBrown99}
P.~A. Durkee, R.~E. Chartier, A.~Brown, W.~J. Trehubenko, S.~D. Rogerson,
  C.~Skupniewicz, and K.~E. Nielsen.
\newblock Composite ship track characteristics.
\newblock \emph{J. Atmos. Sci.}, 57:\penalty0 2542--2553, 2000.

\bibitem[Goren et~al.(2019)Goren, Kazil, Hoffmann, Yamaguchi, and
  Feingold]{Goren_2019}
T.~Goren, J.~Kazil, F.~Hoffmann, T.~Yamaguchi, and G.~Feingold.
\newblock Anthropogenic air pollution delays marine stratocumulus break‐up to
  open‐cells.
\newblock \emph{Geophys. Res. Lett.}, 46:\penalty0 14135--14144, 2019.
\newblock \doi{10.1029/2019gl085412}.

\bibitem[Carslaw et~al.(2013)Carslaw, Lee, Reddingtion, Pringle, Rap, Forster,
  Mann, Spracklen, Woodhouse, Regayre, and Pierce]{CarslawLeeReddingtion13}
K.~S. Carslaw, L.~A. Lee, C.~L. Reddingtion, K.~J. Pringle, A.~Rap, P.~M.
  Forster, G.~W. Mann, D.~V. Spracklen, M.~T. Woodhouse, L.~A. Regayre, and
  J.~R. Pierce.
\newblock Large contribution of natural aerosols to uncertainty in indirect
  forcing.
\newblock \emph{Nature}, 503:\penalty0 67--71, 2013.
\newblock \doi{10.1038/nature12674}.

\bibitem[Fu et~al.(2019)Fu, {Di Girolamo}, Liang, and Zhao]{FuGirolamoLiang19}
D.~Fu, L.~{Di Girolamo}, L.~Liang, and G.~Zhao.
\newblock Regional biases in {MODIS} marine liquid water cloud drop effective
  radius deduced through fusion with {MISR}.
\newblock \emph{J. Geophys. Res. Atmos.}, 124:\penalty0 13182--13196, 2019.
\newblock \doi{10.1029/ 2019JD031063}.

\bibitem[Malavelle et~al.(2017)Malavelle, Haywood, Jones, Gettelman, Clarisse,
  Bauduin, Allan, Karset, Kristj{\'a}nsson, Oreopoulos, Cho, Lee, Bellouin,
  Boucher, Grosvenor, Carslaw, Dhomse, Mann, Schmidt, Coe, Hartley, Dalvi,
  Hill, Johnson, Johnson, Knight, O'Connor, Partridge, Stier, Myhre, Platnick,
  Stephens, Takahashi, and Thordarson]{Malavelle_2017}
F.~F. Malavelle, J.~M. Haywood, A.~Jones, A.~Gettelman, L.~Clarisse,
  S.~Bauduin, R.~P. Allan, I.~H.~H. Karset, J.~E. Kristj{\'a}nsson,
  L.~Oreopoulos, N.~Cho, D.~Lee, N.~Bellouin, O.~Boucher, D.~P. Grosvenor,
  K.~S. Carslaw, S.~Dhomse, G.~W. Mann, A.~Schmidt, H.~Coe, M.~E. Hartley,
  M.~Dalvi, A.~A. Hill, B.~T. Johnson, C.~E. Johnson, J.~R. Knight, F.~M.
  O'Connor, D.~G. Partridge, P.~Stier, G.~Myhre, S.~Platnick, G.~L. Stephens,
  H.~Takahashi, and T.~Thordarson.
\newblock Strong constraints on aerosol--cloud interactions from volcanic
  eruptions.
\newblock \emph{Nature}, 546\penalty0 (7659):\penalty0 485--491, 2017.
\newblock \doi{10.1038/nature22974}.

\bibitem[Wood et~al.(2017)Wood, Ackerman, Rasch, and Wanser]{Wood_17}
R.~Wood, T.~Ackerman, P.~Rasch, and K.~Wanser.
\newblock Could geoengineering research help answer one of the biggest
  questions in climate science?
\newblock \emph{Earth's Future}, 5:\penalty0 659--663, 2017.

\bibitem[Xie and Liu(2013)]{Xie_2013}
Y.~Xie and Y.~Liu.
\newblock A new approach for simultaneously retrieving cloud albedo and cloud
  fraction from surface-based shortwave radiation measurements.
\newblock \emph{Environ. Res. Lett.}, 8\penalty0 (4):\penalty0 044023, 2013.
\newblock \doi{10.1088/1748-9326/8/4/044023}.

\bibitem[Ackerman et~al.(2009)Ackerman, vanZanten, Stevens, Savic-Jovcic,
  Bretherton, Chlond, Golaz, Jiang, Khairoutdinov, Krueger, Lewellen, Lock,
  Moeng, Nakamura, Petters, Snider, Weinbrecht, and
  Zulauf]{AckermanZantenStevens09}
A.~S. Ackerman, M.~C. vanZanten, B.~Stevens, V.~Savic-Jovcic, C.~S. Bretherton,
  A.~Chlond, J.-C. Golaz, H.~Jiang, M.~Khairoutdinov, S.~K. Krueger, D.~C.
  Lewellen, A.~Lock, C.-H. Moeng, K.~Nakamura, M.~D. Petters, J.~R. Snider,
  S.~Weinbrecht, and M.~Zulauf.
\newblock Large-eddy simulations of a drizzling, stratocumulus-topped marine
  boundary layer.
\newblock \emph{Mon. Weather Rev.}, 137\penalty0 (3):\penalty0 1083--1110,
  2009.
\newblock \doi{10.1175/2008mwr2582.1}.

\end{thebibliography}
\end{document}